\documentclass[twocolumn]{aastex631}

\newcommand{\COO}{CO$_2$ }
\newcommand{\cmu}{cm$^{-1}$ }

\usepackage{chemfig}
\usepackage[version=4]{mhchem}

\begin{document}

\title{The survival of aromatic molecules in protoplanetary disks}

\author[0000-0001-6947-7411]{Elettra L. Piacentino}
\affiliation{Harvard-Smithsonian Center for Astrophysics, 60 Garden Street, Cambridge, MA 02138, USA }

\author[0009-0006-4436-9848]{Aurelia Balkanski}
\affiliation{Harvard-Smithsonian Center for Astrophysics, 60 Garden Street, Cambridge, MA 02138, USA }

\author[0000-0002-0150-0125]{Jenny Calahan}
\affiliation{Harvard-Smithsonian Center for Astrophysics, 60 Garden Street, Cambridge, MA 02138, USA }

\author[0009-0009-8813-7442]{Anna Fitzsimmons}
\affiliation{Harvard-Smithsonian Center for Astrophysics, 60 Garden Street, Cambridge, MA 02138, USA }

\author[0000-0003-2761-4312]{Mahesh Rajappan}
\affiliation{Harvard-Smithsonian Center for Astrophysics, 60 Garden Street, Cambridge, MA 02138, USA }

\author[0000-0001-8798-1347]{Karin I. {\"O}berg}
\affiliation{Harvard-Smithsonian Center for Astrophysics, 60 Garden Street, Cambridge, MA 02138, USA }

\begin{abstract}
Aromaticity is a common chemical functionalities in bioactive molecules. In interstellar and circumstellar environments benzene and other small aromatics are considered the precursor for more complex prebiotic molecules and they have shown to potentially have rich ice-phase photochemistry. 
The availability of small organic molecules in prebiotic networks depends on their photostability in astrophysical environments preceding planet formation, particularly during the protoplanetary disk stage, as the disk composition is linked to the chemical make-up of planets and planetesimals.  We study the ultraviolet (UV) photodestruction (120-160 nm) of five aromatic molecules in undiluted ices and, for selected cases, in astrophysically relevant ice matrices (H$_2$O, CO, CO$_2$). For each ice, we measure the destruction cross sections as a function of photon exposure. In undiluted ices, aromatic molecules exhibit substantially lower photodestruction cross sections ($\sigma$$\,<\,$10$^{-19}$ cm$^2$) than aliphatic hydrocarbons, including cyclohexane, ($\sigma$\,=\,2.8-4x10$^{-18}$ cm$^2$). Furthermore, neither substituent nature nor size affects the aromatic stability in pure ices, suggesting that the strong intermolecular interactions among aromatic molecules provide protection against VUV exposure, even with small to mid-sized ring substituents.
In mixed ices, the photodestruction and reactivity of aromatic molecules ($\sigma$=2.5-6.1x10$^{-18}$ cm$^2$) increases by more than an order of magnitude, but are still lower than in the gas-phase. We attribute this to a weaker cage effect and matrix-specific interactions. We use the experimental photodestruction cross sections to estimate the lifetime of aromatic molecules in protoplanetary disks, denileating the disks regions in which aromatic photochemistry is expected to be the most active.

\end{abstract}

\section{Introduction} \label{sec:intro}

Aromaticity is embedded in several molecules of biological importance. From small metabolites to nucleic acids (Deoxyribonucleic acid (DNA) and ribonucleic acid (RNA)), most bioactive molecules contain at least one aromatic feature.
Benzene, along with its functionalized homologues, serves as one of the fundamental building blocks for the synthesis of these bioactive molecules and also serve as model system for the understanding of the synthesis, behavior, and degradation of more complex aromatic species (For a detailed discussion of the relevance of aromatic molecules in biochemistry see e.g., \citet{szatylowicz2021aromaticity, balaban2004aromaticity} and \citet{pozharskii2011heterocycles}).

Large aromatic molecules have long been known to be present in a variety of astrophysical environments in the form of polycyclic aromatic hydrocarbons (PAHs). Benzene, in particular, is often viewed as a fundamental precursor in their bottom-up formation. \citep{lee2019gas,kocheril2025termination}.  The ubiquitous presence of PAH's has made these molecules subjects of many observational and laboratory studies in the past few decades \citep[e.g.][]{leger1984identification,allamandola1989interstellar,tielens1997circumstellar,tielens2013molecular,gudipati2003ApJ...596L.195G,gudipati2006ApJ...638..286G,bouwman2010A&A...511A..33B,cook2015photochemistry,Zeichner2023Sci...382.1411Z}.  However, it was only recently that the presence of small aromatics in space became known, and over the past few years several aromatic molecules have been added to the aromatic interstellar medium (ISM) inventory. Benzene was first tentatively detected in a protoplanetary nebula by \citet{cernicharo2001infrared}, but recent additions to the aromatic inventory now include benzonitrile, the simplest N-bearing aromatic molecule \citep{mcguire2018Sci...359..202M}, as well as 1-cyanonaphthalene, 2-cyanonaphthalene \citep{mcguire2021detection},  and 1-indene \citep{mcguire20182018}.
In disks, the presence of benzene has more recently been confirmed in the inner disk of J160532 M dwarf star \citep{tabone2023rich}.
The aromatic inventory in comets also appears to be significant, with benzene, toluene, phenol, benzonitrile, benzaldehyde, and benzoic acid detected in the coma of comet 67P/Churyumov-Gerasimenko during the ROSETTA mission, indicative of substantial aromatic ice reservoirs in protoplanetary disks \citep{aro67p2019A&A...630A..31S,hanni2022NatCo..13.3639H}.

These observations demonstrate that aromatic molecules are widespread throughout the various stages of star and planet formation. However, it remains unclear whether the aromatic inventory in comets is inherited from the ISM or formed later through disk and parent body chemistry. This uncertainty arises because during disk and planet formation, molecules experience significant thermal variations and exposure to various radiation sources, which can drastically alter the chemical composition, including that of the icy grains. 
When considering thermal variation, models indicate that most grains incorporated into disks did not experience temperatures above water ice desorption \citep{Visser09}. In protoplanetary disks, thermal effects are also expected to act more strongly on volatile species, with water being notably affected only within a few a.u. of the central star \citep{Bergnericeinh2021ApJ...919...45B}.  Given these findings and the high desorption energies of aromatic molecules \citep{piacentino2024characterization}, ice sublimation is likely to have a limited impact on the ice abundance of this class of molecules.
Due to their thermal stability, dissociation via irradiation by high-energy photons or electrons is likely the primary destruction pathway for aromatics. This raises key questions about the extent to which aromatic molecules, when embedded in icy grain mantles, are dissociated and whether this dissociation is primarily destructive or could potentially lead to the formation of more complex compounds.

Several studies have investigated the photolysis and photochemistry of molecules in the ice phase, both as pure compounds and within complex mixtures (e.g. \citet{allamandola1988photochemical,gerakines1996ultraviolet,moore2000ir}). The few laboratory studies that have explored the lifetimes of aromatic molecules exposed to ultraviolet (UV) radiation found significant variability, depending on both the molecular complexity and the overall ice environment. 
In particular, \citet{ruiterkamp2005quantitative} found that the half-life of pure benzene ice exposed to UV in dense cloud conditions is $\sim$\,1 order of magnitude longer compared to when benzene is diluted in water ice (1:5) and $\sim$\,2 order of magnitude longer than when benzene is embedded in a 1:20  CO, or 1:30 \COO ices. A comparable lifetime is also observed in argon mixtures at significantly higher dilutions (1:350 or 1:700), which are intended to approximate gas-phase behavior. This variation in the half-life corresponds to a similar order of magnitude variation in the destruction cross sections.
In other laboratory experiments, \citet{uvvolution2010P&SS...58.1327G} studied the UV destruction cross section of solid purines. While the experimental conditions between the work of \citet{ruiterkamp2005quantitative} and that of \citet{uvvolution2010P&SS...58.1327G} are not exactly identical, an order of magnitude reduction in the destruction cross section values is apparent going from pure benzene to more complex aromatic molecules, such as heteroaromatic and substituted benzenes.
Other studies have investigated the photochemistry of dibenzyl ketone in solution with various large organic solvents, finding that the branching ratio of photochemical pathways is highly sensitive to the nature of the solvent, even at room temperature \citep{rao1986modification}.

These studies highlight that the photostability of aromatic molecules is not yet well understood and depends on a range of factors, including the molecular complexity of the aromatic compound, temperature, and mixture composition. In particular, the effect of aromatic ring substituents on a molecule’s UV resistance has not been investigated in detail. In this study, we explore how the UV resistance of aromatic molecules differ from aliphatic hydrocarbons and how functional group substitution and the nature of the surrounding ice matrix influences aromatic photostability. We pay particular attention to the physical constraints imposed by the ice environment, examining how ice stiffness can modulate photodestruction efficiency in astrophysically relevant ices. In doing so, we assess whether the trends identified by \citet{ruiterkamp2005quantitative} extend across different temperature regimes and to substituted benzenes, where both fragmentation patterns and ice stiffness are expected to vary.
We first test the photodestruction efficiency of undiluted benzene ice at three ice temperatures and compare it to the photodestruction of aliphatic hydrocarbons of different sizes, such as ethane, propane, and cyclohexane as well as the benzene derivatives toluene, ethylbenzene, butylbenzene and benzonitrile (\S \ref{subsec:decaypure}). We next evaluate the photodestruction cross section of the ices of benzene, toluene, and benzonitrile, when mixed with H$_2$O, \COO, and CO. (\S \ref{sec:results})
Finally, we use the experimentally-derived destruction cross sections to estimate the lifetime of small aromatics and of cyclohexane in disk environments where the UV exposure drastically changes both radially and vertically from the central star (\S\ref{sec:astro}).

\section{Experimental Details}

\subsection{Experimental setup}
We study the destruction of small aromatic molecules using the SPACECAT (Surface Processing Apparatus for Chemical Experimentation to Constrain Astrophysical Theories) apparatus which is fully described in \citet{martin2020formation}.

In brief, SPACECAT is an ultra-high vacuum (UHV) spherical chamber that is kept at a pressure of 10$^{-10}$ torr. A helium-cryocooled CsI substrate is mounted in the center of the chamber. Ices are grown by direct condensation from pure or mixed gas-phase samples on to the cooled substrate. Gaseous samples are delivered in close proximity to the substrate surface through a staineless steel tube with a diameter of $\diameter$ 4.8 mm. The flux of the gas mixtures is controlled through a leak valve that is connected to a gas-mixing system that has a base pressure of 1 x 10$^{-4}$ Torr. The temperature of the cryocooled substrate can be controlled between 12 and 300\,K using a temperature controller (LakeShore 335) that has an accuracy of $\pm$\,2\,K  and uncertainty of 0.1\,K. Relative ice composition and absolute ice coverage are monitored in situ using a Fourier transform infrared (IR) spectrometer (Bruker Vertex 70v). A quadrupole mass spectrometer (Pfeiffer QMG 220M1) is also available on the SPACECAT chamber for the  monitoring of gas-phase species present in the chamber. SPACECAT is also equipped with an H$_2$:D$_2$ lamp (Hamamatsu L11798) used to photoprocess the ices (see \S\ref{subsec-vuv}). 
The photon flux is measured for each experiment using a National Institute of Standards and Technology (NIST)-calibrated AXUV-100G photodiode and has an uncertainty of $\sim$ 5\,$\%$ \citep{bergner2019detection}. 

\subsection{Sample Preparation}
Samples are prepared using off-the-shelf reagents having the following specifications
C$_6$H$_6$ (Millipore sigma, 99.8\,$\%$), C$_6$H$_5$CN (Millipore sigma, 99.9\,$\%$), C$_6$H$_5$CH$_3$ (Millipore sigma, 99.8\,$\%$), C$_6$H$_5$CH$_2$CH$_3$ (Millipore sigma, 99.8\,$\%$), C$_6$H$_5$(CH$_2$)$_3$CH$_3$ (Millipore sigma, $>$99\,$\%$), C$_2$H$_6$ (Millipore sigma, 99.9\,$\%$), C$_3$H$_8$ (Millipore sigma, 99.9\,$\%$), CO (Millipore sigma, 99.9\,$\%$), CO$_2$ (Millipore sigma, 99.9\,$\%$), and purified water. Gas-phase reagents were used directly without transfers or purification, while liquid reagents were transferred into resealable glass flasks and purified using several freeze-thaw-pump cycles using liquid nitrogen.

The ices are prepared by direct condensation from gas-phase mixtures. These mixtures are assembled in the SPACECAT gas line by collecting measured amounts of each component in a gas-phase ratio of 1:10. This typically results in an ice-phase ratio of approximately 1:11 to 1:20, depending on the specific experiment (see Table \ref{tab_explist}). After measurement, the gas-phase components are allowed to mix in the gas-line for $\sim$\,5–10 minutes before being delivered and deposited onto the cooled substrate at an approximate rate of $\sim$\,1$\times$10$^{16}$ molecules/min.

\subsection{UV irradiation and the effects of the total ice thickness in diluted ices}\label{subsec-vuv}
\begin{figure*} [h!]
\centering
\includegraphics[width=0.9\textwidth]{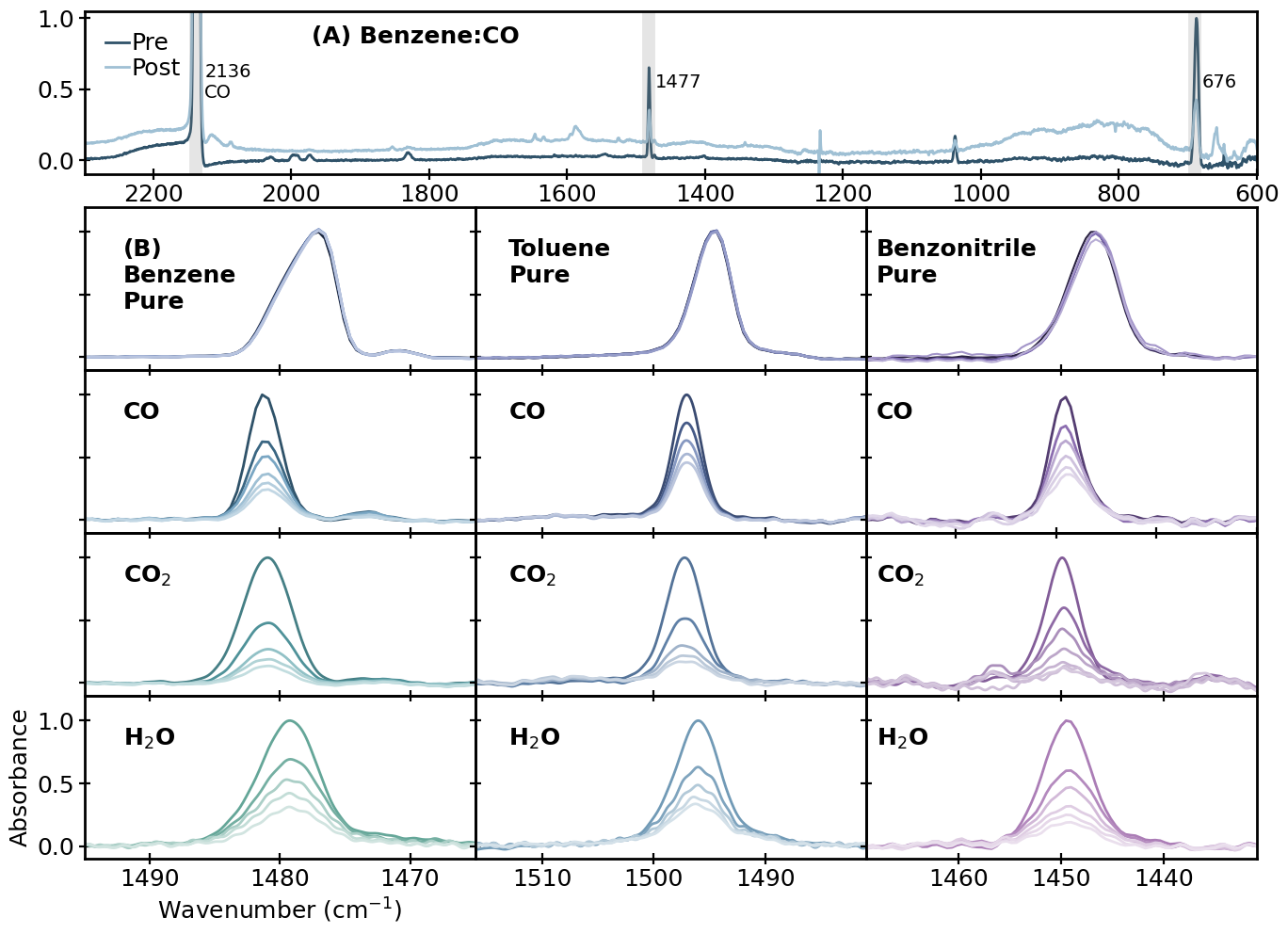}
\caption{(A) Spectral variation of the benzene:CO mixture before and after photon exposure. The dark blue line corresponds to the IR spectra before photon exposure, and the lighter blue corresponds to the IR spectra of the benzene:CO ice after photon exposure.} (B) Spectral variation of the main aromatic bands as a function of photon fluence, with lighter colors representing higher photon fluence. Different colors are used to identify benzene, toluene, and benzonitrile. Each ice was exposed to approximately the same photon fluence of $\sim$7\,$\times$\,10$^{17}$ photons cm$^{-2}$(Table \ref{tab_explist}). Unless explicitly indicated, all subpanels share the same axis scale.
\label{fig: IR main}
\end{figure*}

Aromatic ices destruction and product formation is monitored using the 
area of their most prominent vibrational infrared (IR) band at periodic intervals during photon irradiation of the samples (Fig. \ref{fig: IR main}). The IR bands used along with their respective band strengths are listed for each molecule in Table \ref{table_combined}. An example of the variations that are observed during the experiment for the 600-2000 \cmu spectral range is also shown in panel (A) of Fig. \ref{fig: IR main} for the benzene:CO experiment.

The area of the spectral band taken before ice exposure is converted into monolayers (ML) using Eq. \ref{eqnml} to estimate the initial ice coverage. 
\begin{equation} \label{eqnml}
\rm N = \frac{2.3}{\rm A} \int Abs(\tilde\nu) d\tilde\nu
\end{equation}
where N (molecules cm$^{-2}$) is the column density of the molecule from which the coverage in ML can be obtained in the approximation that 1\,ML\,=\,10$^{15}$ molecules cm$^{-2}$. We note that the use of ML does not necessarily represent the physical thickness of the ice layer in the sample, but is instead used as a convenient unit of column density. A (cm molecule$^{-1}$) is the band strength and $\int$Abs($\tilde\nu$) d$\tilde\nu$ is the area of the IR band in absorbance units.
We note that band strengths for toluene and benzonitrile that are reported in the literature are constrained within a factor of two uncertainty \citep{piacentino2024characterization}, and that those of ethylbenzene and butylbenzene are assumed from that of toluene, therefore carrying a higher uncertainty. However, in our experiments the band strength uncertainty only carries into the estimation of the ice coverage and not into the destruction cross section values as long as the ices are in the optically thin regime. 

\begin{figure} [h!]
\centering
\includegraphics[width=0.95\columnwidth]{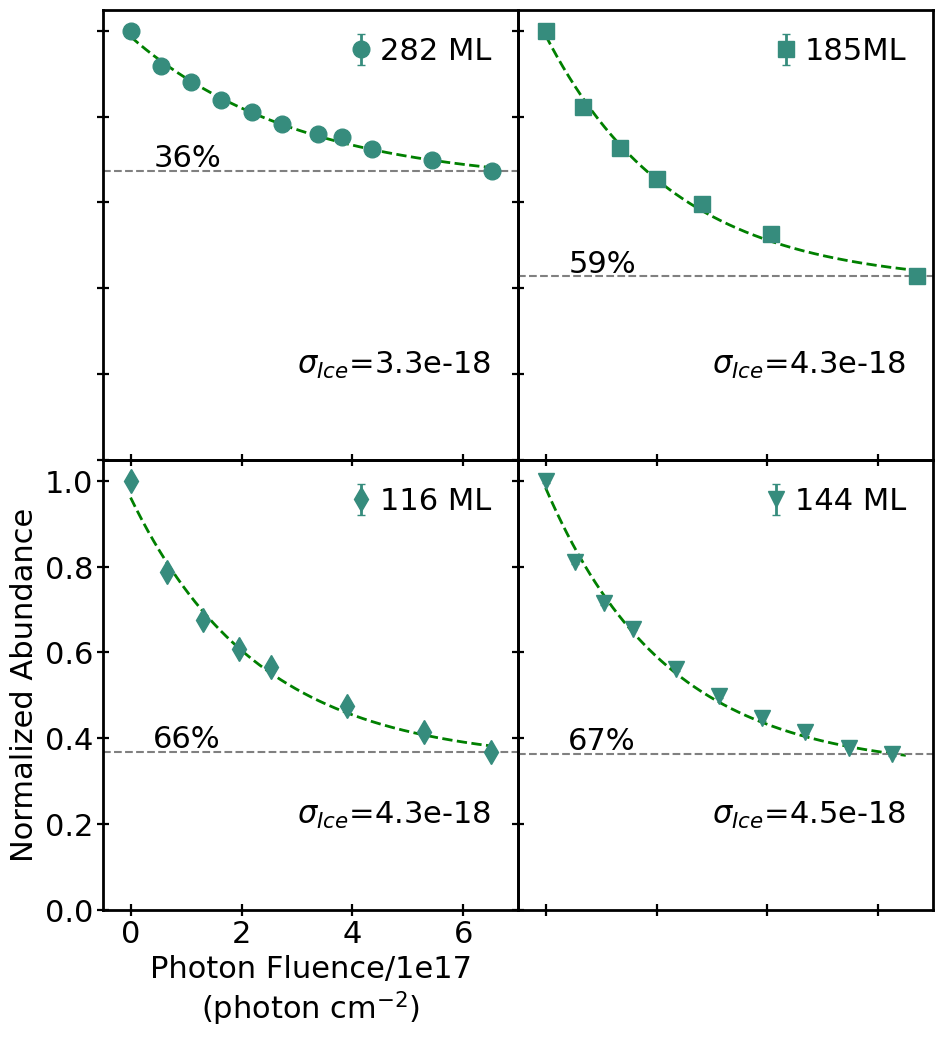}
\caption{Destruction of benzene in benzene:CO ice mixtures with varying photon fluence across four different ice coverages. The fluence for each experiment is reported in Table \ref{tab_explist}. All subpanels share the same axis scale.}
\label{fig:cothickness}
\end{figure}

To ensure that ices are optically thin to the incoming UV radiation, such that the UV flux is approximately constant throughout the ice, we determine the limits of the optically thin regimes empirically using the benzene:CO ice mixture. We both carry out empirical tests and calculate the corresponding theoretical ice thickness cut-off (appendix \ref{app:lamp}). We used one of the matrix experiments since these are the thickest ices, and choose to use CO because it has the highest absorption cross sections values in the 120-160 nm regime among the matrix constituents used (Table \ref{table_combined}), and the most overlap with our lamp emission profile (Fig. \ref{figapp: lamp} in App. \ref{app:lamp}).
In Fig. \ref{fig:cothickness} we compare the destruction profiles and the respective cross sections (see \S\ref{subsec-vuv}) for benzene:CO ices having a ratio of about 1:17 and total ice coverage of 282, 185, 144, and 116 ML. We found that the 185, 144 and 116 ML experiments show very small variation among each other both in terms of destruction yield (59, 66, and 67\,$\%$ respectively) and destruction cross sections (4.3$\times$10$^{-18}$, 4.3$\times$10$^{-18}$ and 4.5$\times$10$^{-18}$ cm$^2$, respectively) while the thickest ice yield values of 36\,$\%$ and 3.3$\times$10$^{-18}$ cm$^2$ for the destruction yield and destruction cross section, respectively. For the purpose of our experiments, we are hence in the optically thin regime when ices are $<$180 ML.
The calculated photon absorption (App. \ref{app:lamp}) by CO for the thickest ice (185\,ML) approaches, 15\,$\%$ at 160\,nm and it is $\sim$\,56\,$\%$ on average in the 120-160\,nm spectral window range (App. \ref{app:lamp}). We note that due to the specific absorption profile of CO in combination with the lamp emission profile, the percent absorption in the 120-160\,nm spectral range is an upper limit. 

\begin{deluxetable*}{c||ccc|ccc}
\tablecolumns{7}
\tablewidth{0.98\textwidth}
\label{table_combined}
\tablecaption{Band strengths values (A) and average absorption cross sections ($\sigma$$_{abs}$)
in the 120-160 nm range for the molecules used in this study.}
\tablehead{
Molecule & Band Position & A/10$^{-18}$ & Ref & $\sigma$$_{abs}$/10$^{-18}$ & Phase & Ref. \\
         & (\cmu)        & (cm molecule$^{-1}$) & & (cm$^2$) & &
}
\startdata
C$_6$H$_6$       & 1477 & 4.8  & $^{[1]}$ & 20   & Gas & $^{[9]}$ \\
C$_6$H$_5$CH$_3$ & 1467 & 3.0  & $^{[2]}$ & 30   & Gas & $^{[10]}$ \\
C$_6$H$_5$CH$_2$CH$_3$ & 1495 & 3.0$^*$  & $^{[2]}$ & -  & - & - \\
C$_6$H$_5$(CH$_2$)$_2$CH$_3$ & 1496 & 3.0$^*$  & $^{[2]}$ & -   & - & - \\
C$_6$H$_5$CN     & 1491 & 1.9  & $^{[2]}$ & -    & -   & - \\
C$_2$H$_6$       & 817  & 1.99 & $^{[3]}$ & 10   & Gas & $^{[11]}$ \\
C$_3$H$_8$       & 1368 & 0.91 & $^{[4]}$ & -    & -   & - \\
C$_6$H$_{12}$ & 2927 & 53 & $^{[5]}$ & 39 & Gas & $^{[12]}$ \\
CO               & 2136 & 11   & $^{[6]}$ & 4.7  & Ice & $^{[13]}$ \\
H$_2$O           & 3280 & 200  & $^{[7]}$ & 3.4  & Ice & $^{[13]}$ \\
CO$_2$           & 2340 & 110  & $^{[8]}$ & 0.6  & Ice & $^{[14]}$ \\
\enddata

\tablenotetext{}{$^*$ Estimated from that of Toluene$^{[2]}$.
$^{[1]}$\citet{hudson2022infrared},
$^{[2]}$\citet{piacentino2024characterization}, 
$^{[3]}$\citet{hudson2014}, 
$^{[4]}$\citet{hudson2021},
$^{[5]}$\cite{Hendecourt1986time}},  
$^{[6]}$\citet{Gerakines2023MNRAS.522.3145G}, 
$^{[7]}$\citet{Gerakines95}, 
$^{[8]}$\citet{Bouilloud2015MNRAS.451.2145B}, $^{[9]}$\cite{Capalbo2016Icar..265...95C}, 
$^{[10]}$\cite{serralheiro_toluene_2015},$^{[11]}$\citet{Chen2004JQSRT..85..195C},$^{[12]}$\cite{bandeira2025cyclohexane}, 
$^{[13]}$\citet{pCruz-Diaz2014AA...562A.119C},
$^{[14]}$\citet{nonpCruz-Diaz2014AA...562A.120C}
\end{deluxetable*}

\subsection{Photodissociation cross section calculations}\label{sec:kinetics}

The photodestruction rate of molecules in the ice phase differs from the gas-phase photodissociation cross sections due to additional processes that are unique to the condensed phase. In ices, ultraviolet (UV) irradiation can trigger a range of simultaneous chemical reactions, making the study of photodestruction in this phase significantly more complex. 
To simplify this, we focus on a model that looks only at the dissociation of a sample molecule, XY, into radicals, the subsequent recombination of these radicals to form XY again, plus a loss term.

Upon absorption of VUV photons, the parent molecule fragments to form radical species (Eq. \ref{eq: chemreac1}). This process can be described by the dissociation cross section, $\sigma_{Ds}$, relative to the photon fluence ($\phi$).

\begin{equation} \label{eq: chemreac1}
\rm XY_{(Ice)} + \textit{hv} \overset{\sigma_{Ds}} \longrightarrow X^._{(Ice)} + Y^._{(Ice)}
\end{equation}

These radicals (X$^.$$_{(Ice)}$ + Y$^.$$_{(Ice)})$ can recombine either directly (Eq. \ref{eq: chemreac2}) with each other or after diffusing through the ice (Eq. \ref{eq: chemreac3}). The efficiency of these processes can be described respectively by the cross sections $\sigma_{Rc}$ and $\sigma_{Df+Rc}$. Finally, any other chemical reaction yielding the loss of XY is described by $\sigma_{Rx}$ (Eq. \ref{eq: chemreac4}).

\begin{equation} \label{eq: chemreac2}
\rm X^._{(Ice)} +  Y^._{(Ice)} \overset{\sigma_{Rc}}{\longrightarrow} XY_{(Ice)}
\end{equation}

\begin{equation} \label{eq: chemreac3}
\rm X^._{(Ice)} +  Y^._{(Ice)} \overset{\sigma_{Df+ Rc}}{\longrightarrow} XY_{(Ice)}
\end{equation}

\begin{equation} \label{eq: chemreac4}
\rm X^._{(Ice)},  Y^._{(Ice)} \overset{\sigma_{Rx}}{\longrightarrow} Products
\end{equation}

From this set of chemical equations we can derive a kinetic model for the disappearance of XY from the ice:

\begin{equation} \label{eq: chemreac5}
\rm \frac{d[XY]}{d\phi} =-\sigma_{Ds}[XY]+\sigma_{Rc}[XY] +\sigma_{Df+Rc}[XY]
\end{equation}

Which in the steady state approximation becomes,

\begin{equation} \label{eq: chemreac6}
\rm \frac{d[XY]}{d\phi} =-\sigma_{Ds}[XY]+(\frac{\sigma_{Ds}(\sigma_{Rc}+\sigma_{Df+Rc})}{\sigma_{Rc}+\sigma_{Df+Rc}+\sigma_{Rx}})[XY]
\end{equation}

This admit a solution in the form of:

\begin{equation} \label{eq:decay}
   \rm Ab_{[XY]}(\phi) = De^{-\sigma_{Ice}\phi} +C
\end{equation}

where Ab$_{[XY]}$ indicates the rate of disappearance of the XY molecule from the ice as a function of photon fluence, $\phi$ (cm$^{-2}$). The factor D describes the initial condition of the system, while the factor C is the steady-state abundance that accounts for the balance between destruction, recombination and other chemical reactions. Finally, $\sigma_{Ice}$ (cm$^2$) represents the total cross section for the disappearance of molecule XY from the ice, and is a linear combination of the cross sections for all the processes occurring within the ice (Eq. \ref{eq: chemreac7}).  

\begin{equation} \label{eq: chemreac7}
\sigma_{Ice}= \sigma_{Ds}- (\frac{\sigma_{Rc}+\sigma_{Df+Rc}}{\sigma_{Rc}+\sigma_{Df+Rc}+\sigma_{Rx}})
\end{equation}

\subsection{Error estimation} \label{subsec:unce}

There are several possible sources of uncertainty that can impact the extracted photodissociation cross sections.
The experimental conditions under which these values are derived have uncertainties both in terms of absolute ice temperature ($\pm$\,2\,K), total ice coverage (20-50\,$\%$), and aromatic mixing ratio (20-50\,$\%$) \citep{bergner2019detection,piacentino2024characterization}.
However, none of these influence the derived photodissociation cross section directly as long as the ices are optically thin. Instead the main sources of error are dominated by the the 5\,$\%$ uncertainty on the photon flux, followed by the  experimental variance, and the uncertainty resulting from the data fit using Eq. \ref{eq:decay}. Upon standard error propagation, we find the uncertainty of the destruction cross sections to be $\sim$\,11$\%$. 

\begin{deluxetable*}{c||cccccc||ccc}
\tablecolumns{7}
\tablewidth{\textwidth}
\label{tab_explist}
\tablecaption{List of experiments and associated destruction cross-sections ($\sigma_{Ice}$). Ice coverages Arom. Cov. and Matrix cov. are reported in monolayers (ML). Max Fluence corresponds to the photon fluence at which the ice is exposed by the end of the experiment and the temperature (T) in K indicates the temperature at which the ice was irradiated. 
The values reported for $\sigma_{Ice}$ and C are derived from the fit using Eq. \ref{eq:decay} for each experimental scenario. The uncertainty in the $\sigma_{Ice}$ values is 11$\%$ (see section \ref{subsec:unce}). $\sigma_{Gas}$ refers to the gas-phase destruction cross-section reported in the literature.}
\tablehead{Molecule&  Matrix& Arom. Cov. & Matrix cov.&   Max  fluence&  T& Ratio & $\sigma_{Ice}$ & C & $\sigma_{Gas}$ \\
 & & (ML)&(ML)& (cm$^{-2}$/10$^{18}$)& (K)& &(cm$^{2}$)/10$^{-18}$ &&(cm$^{2}$)/10$^{-18}$}
\startdata
C$_2$H$_6$ & - & 55 & - & 0.58 & 10 & - & 4.0   & 0.45 & 20$^{\dagger [1]}$\\ 
C$_3$H$_8$ & - & 40   & - & 0.63& 10& -& 3.3  & 0.51 &-\\
C$_6$H$_{12}$ & - & 20   & - & 0.58& 10& -& 3.7  & 0.45 &-\\
\hline
\hline
Benzene  &  -       &  56      &-      &  0.62     &  10     & -& $<$0.1& -    & 30$^{\dagger [2]}$\\
         &  -       &  9        &-      &  0.66     &  10     & - &$<$0.1& -    & -\\
         &  -       &  60      &-      &  0.65     &  100    &-&$<$0.1& -    & -\\\
         &  -       &  45      &-      &  0.65     &  120    &- &$<$0.1& -    & -\\\
         &  CO      &  16      &266    &  0.65     &  10     & 1:16 &3.3 & 0.64&-\\
         &  CO      &   10       &175    & 0.67      & 10      & 1:17.5 &4.3& 0.41&- \\
         &  CO      &   7       &137    & 0.63      & 10      &1:19.6 & 4.5 & 0.33&-\\
         &  CO      &   7       &109   & 0.65       & 10      & 1:16 & 4.3   & 0.34 &- \\
         & CO$_2$    & 6        & 81      & 0.61      & 10     & 1:13.5 & 5.8   & 0.11 &-\\
         &H$_2$O    & 10        & 109  & 0.6       & 10     & 1:11& 3.9   & 0.26 &-\\ [0.15cm]
 Toluene & -        & 44       &-      & 0.67     & 10     & - & $<$0.1  & -  &-\\
         & CO       & 8         &147    & 0.63     & 10     & 1:18& 4.7   & 0.42 &-\\
         & CO$_2$   & 6        &93       & 0.61      & 10     & 1:15.5& 5.0   & 0.1  &-\\
         & H$_2$O   & 9          &126    & 0.6       & 10     & 1:14& 5.8   & 0.34 &-\\[0.15cm]
 Benzonitrile& -    & 26       &-      & 0.65      & 10     && $<$0.1 & -   &-\\
         & CO       & 6         &84   & 0.65      & 10     & 1:14& 2.5   & 0.49 &- \\
         & CO$_2$   & 5        &94      & 0.44      & 10     & 1:19& 5.9   & 0.08 &-\\
         & H$_2$O   & 12        &133    & 0.65      & 10     & 1:11& 4.3   & 0.28 &-\\
 Ethylbenzene& -    & 29       &-      & 0.62      & 10     &-& $<$0.1 & -   &-\\
 Butylbenzene& -    & 29       &-      & 0.63      & 10     &-& $<$0.1 & -   &-\\
\enddata
\tablenotetext{\dagger}{100-140 nm. $^{[1]}$\citet{Rennie1998CP....229..107R},$^{[2]}$ \citet{database2017}}
\end{deluxetable*}

\section{Experimental Results}\label{sec:results}
A comprehensive list of the dissociation experiments is presented in Table \ref{tab_explist}.
First, we study the dissociation efficiency of pure organic ices (\S\ref{subsec:decaypure}) by monitoring the most intense vibrational modes as a function of photon fluence for each molecule. This includes benzene, toluene, ethylbenzene, butylbenzene, benzonitrile, and cyclohexane, as well as the smaller hydrocarbons ethane and propane. The variety of molecules studied under pure ice conditions helps us understand the extent to which molecular size and functionality influence their ability to resist UV exposure.
Next, we present on the effects that the presence of different matrices have on the dissociation cross section of a subset of the aromatic molecules studied (\S\ref{subsec:decaydil}). 

\subsection{Photodestruction of pure organic ices}\label{subsec:decaypure}

\begin{figure*} [h!]
\centering
\includegraphics[width=0.9\textwidth]{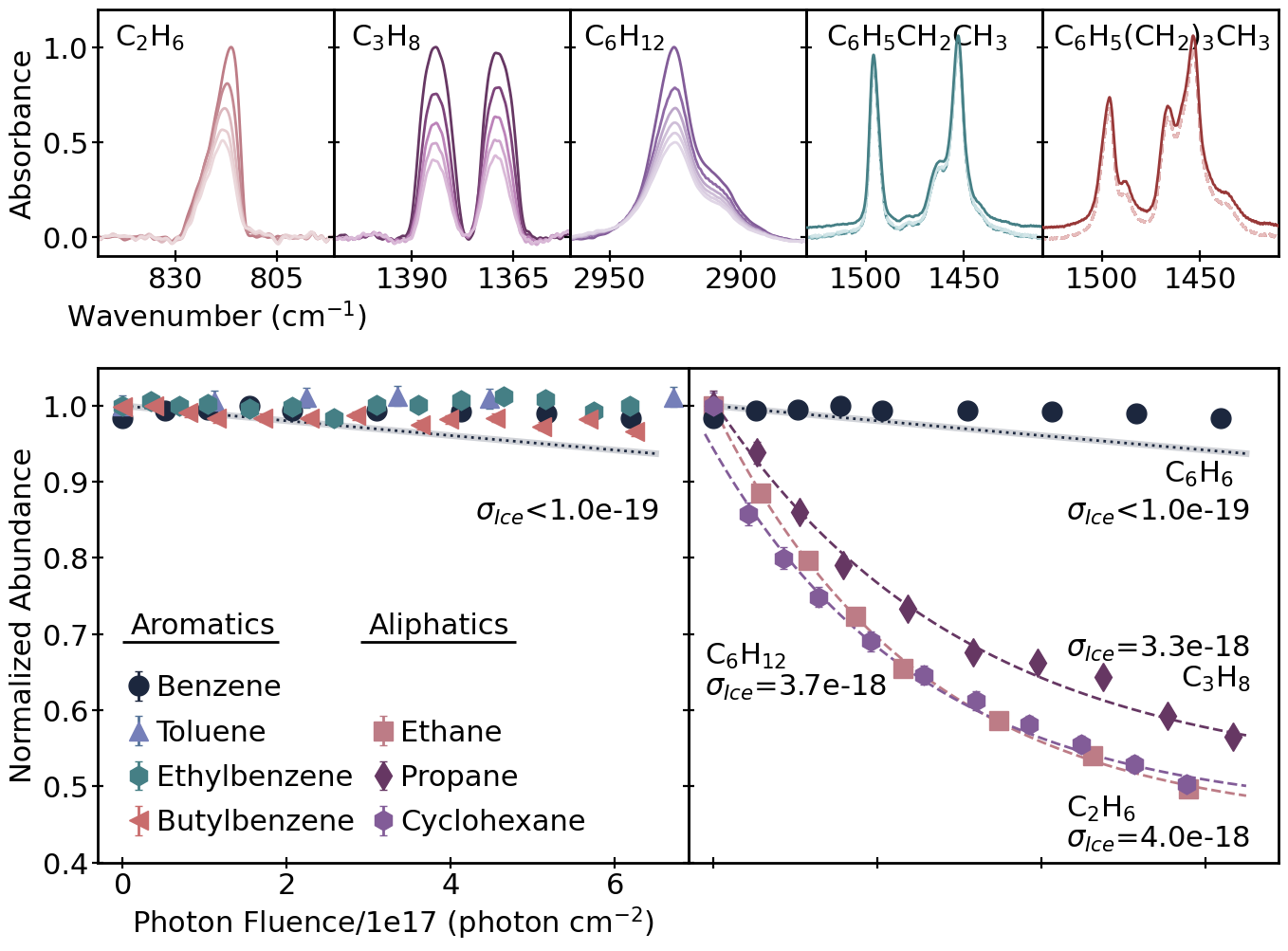}
\caption{Top panels: Variation with fluence of the IR band used for integration of, from left to right, ethane, propane, cyclohexane, ethylbenzene, and butylbenzene. Lighter colors represent higher photon fluence. Given the lack of photodissociation in the cases of ethylbenzene and butylbenzene the trace at fluence\,=\,0 is offset for clarity. The fluence for each experiment is reported in Table \ref{tab_explist}. Bottom left panel: Destruction profile and cross-sections of the undiluted aromatic molecules ices specifically, benzene, toluene, ethylbenzene, and butylbenzene, with a destruction cross-section of $\sigma$$_{ice}$\,$<$\,1\,\,$\times$\,10$^{-19}$ representing an upper limit for all aromatic molecules. Bottom right panel: Destruction profile and cross-sections of undiluted benzene, and of the aliphatic molecules cyclohexane, propane, and ethane, with the destruction cross-section of benzene representing an upper limit.}
\label{fig:allpure}
\end{figure*}

\begin{figure*} [h!]
\centering
\includegraphics[width=0.6\textwidth]{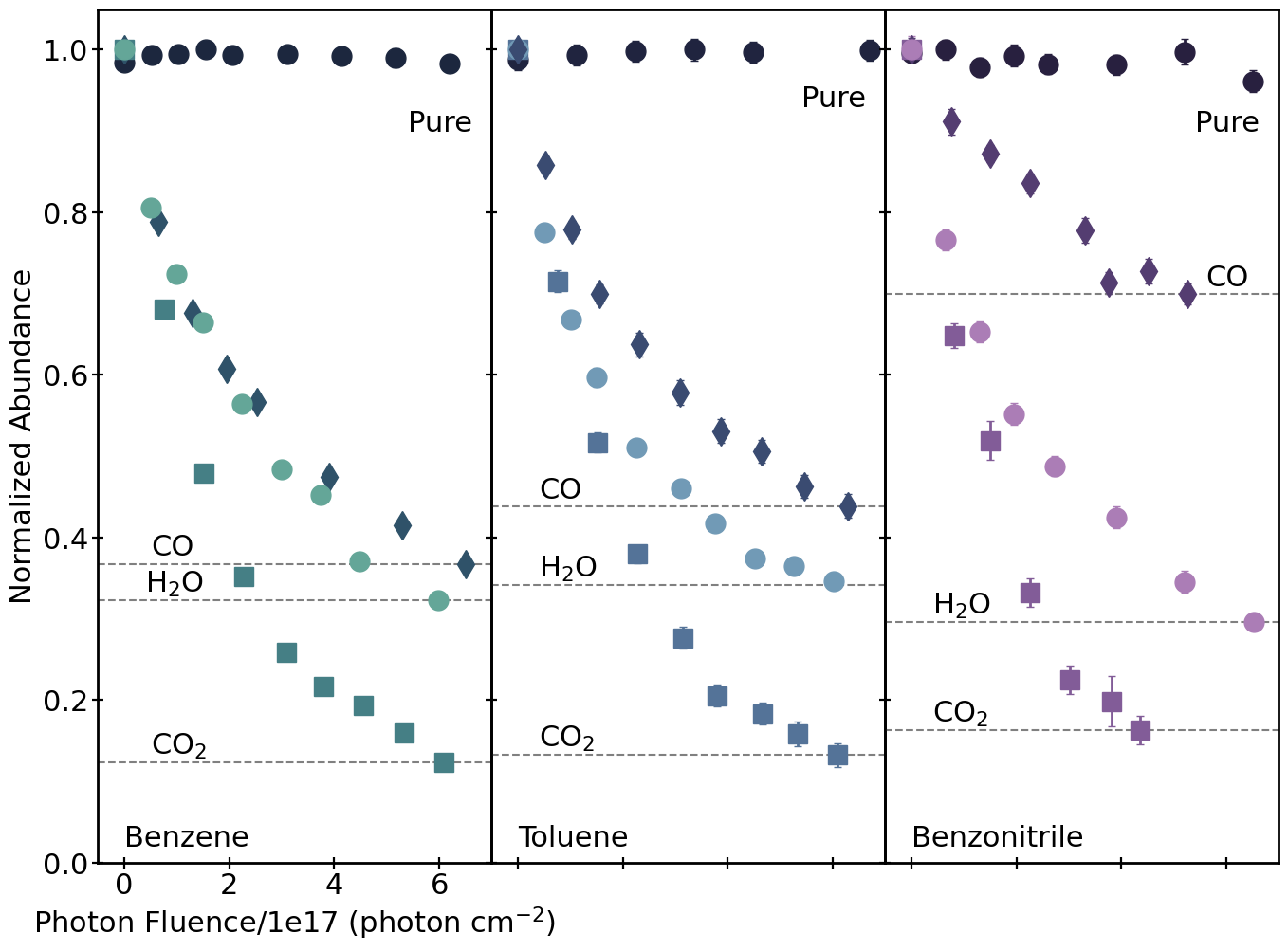}
\caption{Variation in the 
area of the aromatic IR bands as a function of photon fluence for undiluted and mixed ices of benzene (left), toluene (middle), and benzonitrile (right). The fluence for each experiment is reported in Table \ref{tab_explist}. All subpanels share the same axis scale.}
\label{fig:multipanellall}
\end{figure*}

Fig. \ref{fig: IR main} shows the variation in the IR bands for the benzene, toluene, and benzonitrile experimental series. In the top panel of Fig. \ref{fig:allpure}, the changes in the IR bands are shown as a function of photon fluence for the aliphatic hydrocarbons, as well as for ethylbenzene and butylbenzene. The bottom panels display the decay curves, as a function of fluence and relative to that of benzene, for the aromatic species (left) and the aliphatic species (right).
Regardless of the functionality present on the ring, aromatic ices do not display any measurable destruction, we attribute the slight variation in 
band area visible in Fig. \ref{fig:multipanellall} to a variation in band strength with irradiation. For undiluted aromatic ices, our experiments estimate the destruction cross section to be $<$\,1x10$^{-19}$\,cm$^2$, corresponding to $<$\,5\,$\%$ consumption. This value is significantly lower than the destruction cross section reported for gas-phase benzene which is 30×10$^{-18}$\,cm$^2$ \citep{database2017}. 

\begin{figure} [h!]
\centering
\includegraphics[width=0.95\columnwidth]{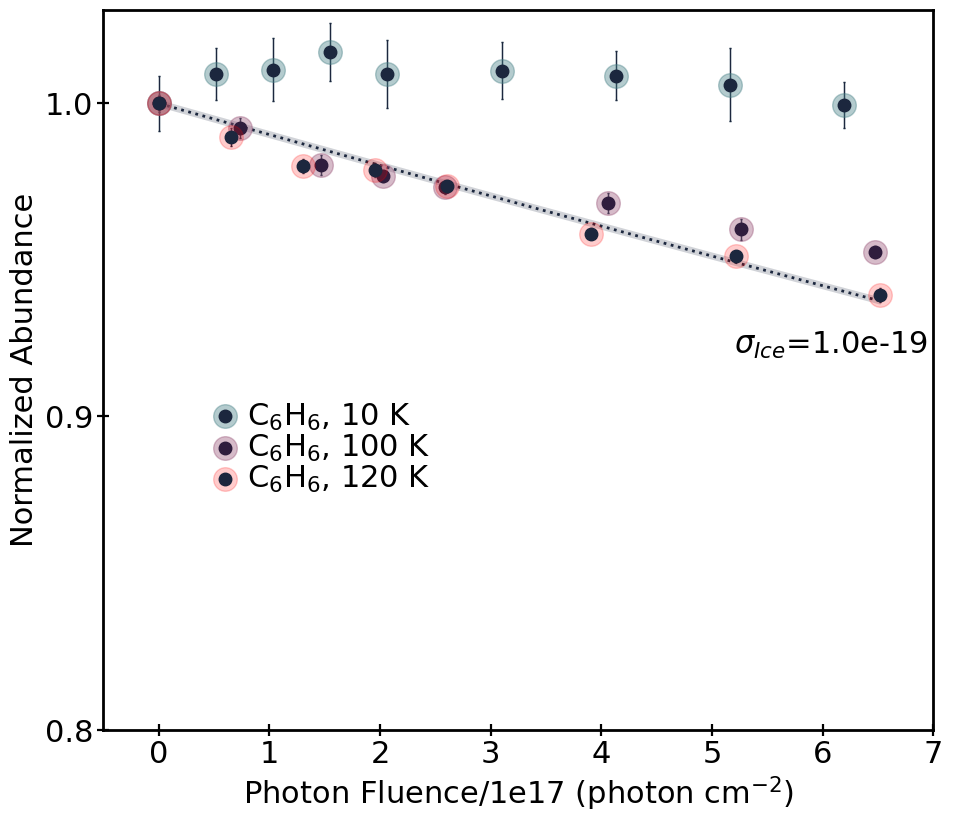}
\caption{Variation in the abundance of benzene in pure ices as a function of photon fluence at three temperatures. A fit using a value of $\sigma$=1x10$^{-19}$ cm$^2$ is shown for reference.}
\label{fig: benzenetemperature}
\end{figure}

The difference between gas-phase and ice-phase destruction cross-sections for pure benzene suggests that limited radical mobility in the ice, which is temperature-dependent, may be a contributing factor. We therefore test whether ice temperature is a significant parameter for the undiluted benzene ices by running experiments similar to our 10 K fiducial at 100 and 120\,K (Fig. \ref{fig: benzenetemperature}). The temperatures are chosen to be high enough to induce a measurable effect in the destruction cross section, but low enough to not cause desorption of the benzene ice, which begins at $\sim$\,138\,K \citep{piacentino2024characterization}. A slightly temperature dependency of the destruction of benzene can be detected, however even at the highest temperature of 120\,K, the destruction remains modest, with $<$10$\%$ of the initial benzene being consumed.  

We also test whether aromaticity as well as aliphatic molecular size are important parameters affecting the destruction cross-section by comparing the destruction of undiluted benzene ice with that of undiluted ices of cyclohexane, ethane, and propane.
The bottom right panel of Fig. \ref{fig:allpure} shows the destruction of benzene, cyclohexane, propane, and ethane, as a function of photon fluence, a summary of the fit parameters and resulting destruction cross sections is presented in Table \ref{tab_explist}.
The smaller organic molecules, ethane and propane, are much less resistant to VUV in their pure ice form than benzene yielding 49\,$\%$ and 55\,$\%$ photo destruction, corresponding to cross sections of  3.3×10$^{-18}$ and 4.0×10$^{-18}$\,cm$^2$ respectively. This is also the case for the larger aliphatic hydrocarbon, cyclohexane, for which the largest destruction cross section value of 3.7×10$^{-18}$\,cm$^2$ and destruction yield of 45 $\%$ are obtained.
The difference between benzene and aliphatic hydrocarbons is surprising, as in the gas phase these molecules are reported to have similar destruction cross-sections, with values of 30×10$^{-18}$ cm$^2$ for benzene and 20×10$^{-18}$ cm$^2$ for ethane. However, both for ethane and benzene, there is a decrease in the destruction cross-sections going from gas to ice, with ethane showing a one order of magnitude decrease and benzene a two order of magnitude decrease (Table \ref{tab_explist}).

\subsection{Photodestruction of diluted aromatic ices}\label{subsec:decaydil}
\begin{figure*} [h!]
\centering
\includegraphics[width=0.8\textwidth]{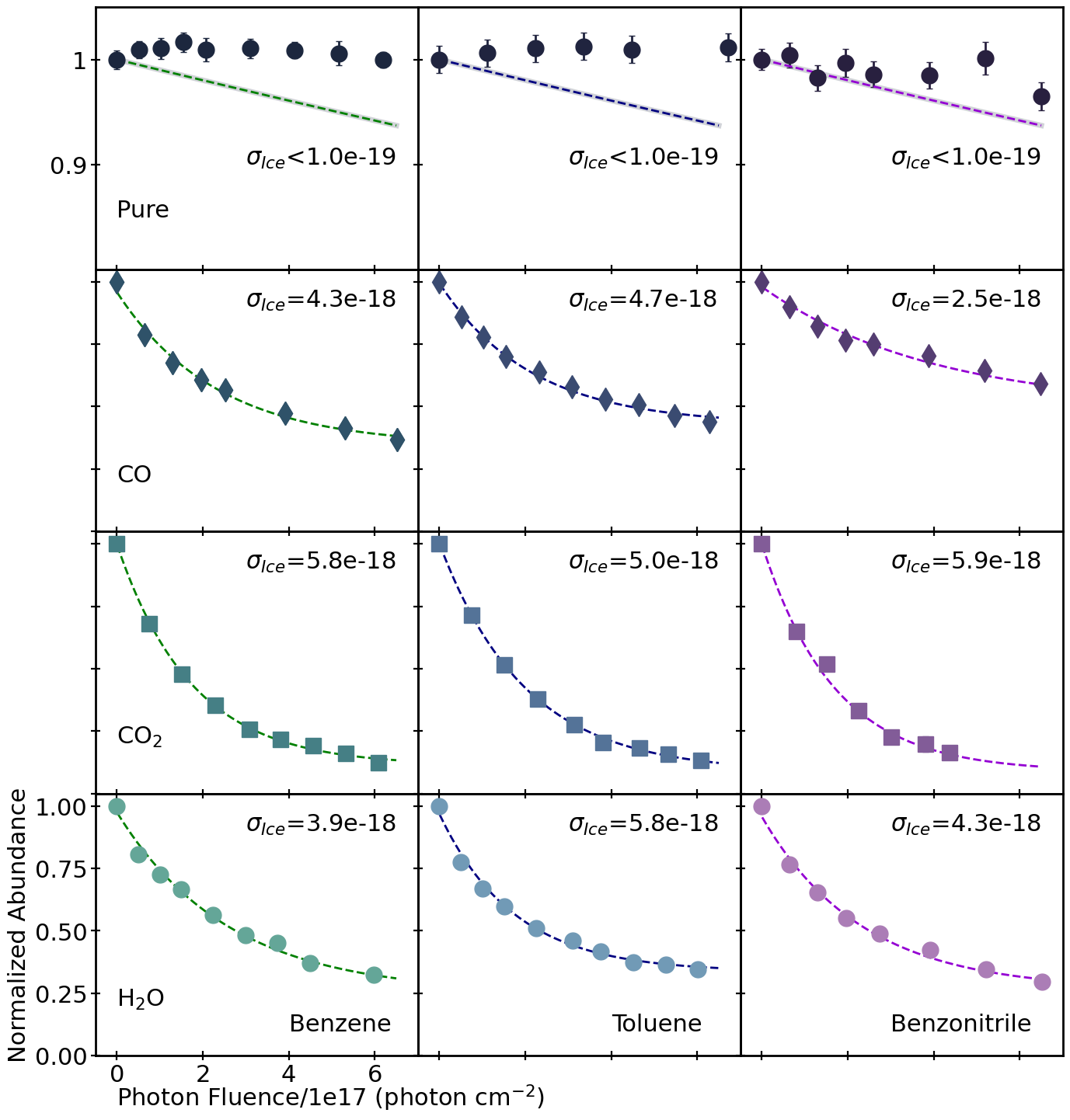}
\caption{First-order fit and destruction cross section ($\sigma_{Ice}$) for the disappearance curve in each experimental scenario. For pure ices we report upper limits on the photodestruction cross sections. The fluence for each experiment is reported in Table \ref{tab_explist}. Unless explicitly indicated, subpanels share the same axis scale. A fit using a value of $\sigma$=1x10$^{-19}$ cm$^2$ is shown for reference for pure ices in the top three panels.}
\label{fig:fitall}
\end{figure*}

The effect of matrix composition on the photodestruction of aromatic molecules is explored by examining the aromatic destruction efficiency in different ice environments (CO, H$_2$O, and \COO) that are expected to be common in space. Fig. \ref{fig:fitall} shows the variation of the
IR spectral band area for each experimental scenario as a function of photon fluence.

In the presence of CO, we find that the fraction of destroyed aromatics increases by an order of magnitude in all cases compared to pure ices. However, the destruction cross-section of benzene:CO ice remains an order of magnitude lower than the gas-phase destruction cross-section for benzene.
The steady state destruction yields varies across the three molecules, ranging from 51\,$\%$ to 65\,$\%$, and the cross sections differ by up to 40\,$\%$, with benzene having the highest and benzonitrile the lowest cross section.

The photodestruction efficiency increases further when the aromatics are instead embedded in a H$_2$O matrix yielding destruction cross sections between 3.9-6.1$\times$10$^{-18}$ cm$^2$, a 74-66\,$\%$ disappearance of the icy aromatics by the time steady state is reached. The presence of \COO as a diluent increases the destruction yield even more compared to the water mixtures, yielding 89-92\,$\%$ destruction respectively for benzene, toluene and benzonitrile, destruction cross section values slightly decrease to 5-5.8$\times$10$^{-18}$ cm$^2$.

In conclusion and with the caveat that the difference is small enough that it may be explained by experimental variation, we find a slight dependence of the destruction cross section on the specific aromatic and matrix constituents. We find a correlation between the matrix constituent and the steady-state abundance (C values in Table \ref{tab_explist}), which are the lowest for CO$_2$ and the highest for CO, for all three aromatic molecules.
The destruction cross section values obtained in the three matrix constituents, follow different trend for each aromatic molecules. For benzene, we find a trend of increasing destruction cross section in the order H$_2$O$<$CO$<$CO$_2$, which agrees with the results of \citet{ruiterkamp2005quantitative}. However, for toluene and benzonitrile, we observe different trends: CO$<$CO$_2$$<$H$_2$O for toluene and CO$<$H$_2$O$<$CO$_2$ for benzonitrile. This inconsistencies suggest that different reactions are contributing to the measured $\sigma_{Ice}$ in each experimental scenario. 

\section{Discussion} \label{sec:disc}

\subsection{Photodissociation of undiluted organic ices}\label{subsecdis:undiluted}

In our experiments, we determine the photodestruction cross section in the ice phase using Eq. \ref{eq:decay}, as described in \S\ref{sec:kinetics}. This differs from gas-phase photodissociation cross sections due to the presence of additional competing chemical processes in the condensed phase.
The experimental photodestruction cross section of aromatic compound ices differs from that of aliphatic hydrocarbons by several orders of magnitude. As discussed above the gas-phase absorption cross sections of aromatics are similar to those of ethane and cyclohexane (Table \ref{table_combined}), making the observed difference surprising (Fig. \ref{fig:cothickness}). Furthermore, \citet{pCruz-Diaz2014AA...562A.119C} and \citet{nonpCruz-Diaz2014AA...562A.120C} report average absorption cross section of a few molecules in the solid state. 
In the case of methane and methanol (i.e. the most ``hydrocarbon-like" molecules included in the study by \citet{pCruz-Diaz2014AA...562A.119C, nonpCruz-Diaz2014AA...562A.120C}), the ice-phase 120-160\,nm absorption cross sections are within a factor of two of the gas-phase counterparts.
Consequently, it is not likely that photon absorption variability explains the order of magnitude decrease in undiluted aromatic ices compared to their aliphatic counterparts.

We note that our undiluted aromatic destruction cross sections, consistent with those reported by \citet{ruiterkamp2005quantitative} for benzene ice, are one to two orders of magnitude lower than the destruction cross sections for gas-phase benzene \citep[$\sim$3$\times$10$^{-17}$ cm$^2$][]{database2017}, suggesting that the answer lies in the behavior of organics in the ice phase. Perhaps the most obvious impact of being embedded in an ice is its limitation on movement of reactants and products alike. 

The so-called cage effect refers to the phenomenon where radicals formed via photolysis become temporarily trapped by surrounding unfragmented or solvent molecules, effectively forming a "cage" that restricts their motion \citep{mcneill2012organics}. Within this cage, the photolysis fragments can undergo one of two competing processes. They may recombine to reform the parent molecule or diffuse away and react with other fragments formed in the lattice. The branching ratio between recombination and diffusion depends on several factors, including the size and nature of the fragments, the ice composition, and the temperature \citep{braden2001solvent} with recombination being favored in stiffer matrices and at lower temperatures \citep{Oberg16}. In the case of 10 K aromatic ices, all these factors contribute to the very efficient recombination we observe.

Radical mobility is most favorable for smaller fragments and with higher temperature. Photo-fragmentation of benzene produces a significant fraction of benzyl radicals \citep{ni2007photodissociation}, which is much larger than the C$_2$H$_5$ radical formed in the photolysis of ethane \citep{price2003photodissociation} explaining the difference in radical diffusion efficiency between aromatic and small aliphatic ice destruction. Our results show that this trend can be extended to other aromatic molecules, as no fragmentation products from the side functional groups have been observed in the IR spectra of pure benzonitrile, toluene, ethylbenzene, and butylbenzene ices (Appendix \ref{app:productsfigandtable}). However, differences in radical size alone do not fully explain the contrasting behavior observed between benzene and cyclohexane. It has been shown that the VUV photolysis of cyclohexane vapor produces a variety of radicals including methyl, allyl, and predominantly cyclohexyl radicals \citep{sevilla1970radical,pilling2012formation,ramphal2021photodissociation}. The cyclohexyl radical is similar in size to the benzyl radical and would therefore be expected to exhibit comparable mobility in ice. Yet, unlike the benzyl radical, which retains a rigid aromatic structure, the cyclohexyl radical can undergo ring-opening reactions to form more flexible species such as hexene \citep{pilling2012formation,ramphal2021photodissociation}. These structural rearrangements increase conformational flexibility, which enhance radical mobility and reactivity in cyclohexane ices compared to benzene ice in which aromaticity is preserved in the benzyl radical.

Aromaticity does not only inhibits the formation of small, mobile fragments and ring-opening products but also contributes to the increased stiffness of benzene ice compared to aliphatic cyclohexane ice. In benzene, the ice matrix is composed of unfragmented aromatic molecules that can form strong $\pi$–$\pi$ interactions when clustered \citep{Mahadevi2010JChPh.133p4308M}, contributing to a significantly stiffer structure than that formed by aliphatic molecules like cyclohexane \citep{cabaleiro2018nature}. Because aromatic molecules exhibit similar stacking interactions, comparable increases in ice rigidity can be expected across aromatic ices more generally. Such enhanced rigidity is consistent with reported interaction energies and intermolecular distances, which show stronger binding and closer packing in benzene dimers compared to those of cyclohexane \citep{kim2011different, cabaleiro2017sigma}, Specifically, the benzene–benzene interaction distance (3.54 Å) is considerably shorter than that of cyclohexane–cyclohexane (4.64 Å) \citep{cabaleiro2017sigma}, indicating tighter packing in the benzene matrix. As a result, fragment mobility is more restricted in benzene ice, whereas in the more loosely packed cyclohexane ice, fragments are freer to diffuse and react.
The cage effect is particularly evident in ethylbenzene and butylbenzene, the molecules bearing the largest substituents. In these instances, the side chains would be expected to fragment more readily, resulting in a higher fragmentation yield than in molecules with shorter substituents, such as toluene. However, this is not observed, indicating that the cage effect, arising from the stacking interactions between the aromatic rings, effectively promotes efficient recombination. 
Finally, the increase, albeit small, of benzene photodestruction with temperature (Fig. \ref{fig: benzenetemperature}) further supports recombination as a cause of the lack of photodissociation of aromatic ices as higher ice temperatures promote ice diffusion.

In conclusion, and in agreement with previous findings \citep[e.g.][]{ruiterkamp2005quantitative}, aromaticity appears to play a key protective role in the ice phase under VUV irradiation, with substitution having no significant effect on disrupting the aromatic cage up to molecules the size of butylbenzene. The delocalized $\pi$-electron system in benzene and in other aromatic molecules, not only stabilizes the molecule against complete fragmentation but also contributes to the formation of a rigid, tightly packed ice matrix through strong intermolecular interactions. This structural rigidity increase the cage effect and reduces fragment mobility and reactivity, resulting in lower destruction cross sections compared to aliphatic analogs. This suggests that aromaticity serves as an intrinsic stabilizing factor against energetic processing in astrophysical ice analogs, even in the presence of single, relatively large ring substituents.

We note that substituents larger than those studied, or possibly multiple substituents on the aromatic ring, could still affect the observed UV resistance by disrupting the $\pi$-$\pi$ interactions in the ice. Therefore, this should be experimentally investigated to determine the limits of the UV-protective effect of aromaticity. Additionally, previous studies have shown that ice morphology can influence the interaction of organics with UV radiation, as the formation of crystals can disrupt the UV behavior of the ices \citep{lignell2015mixing}, potentially altering local UV resistance in regions where crystals form. Given this, dedicated experiments are needed to investigate the role of microcrystallinity in the UV resistance of aromatic ice.

\subsection{Diluted aromatic ices}\label{subsecdis:diluted}

The sharp increase in the photodestruction cross section observed in the presence of a matrix compared to the undiluted aromatic ices further indicates that the aromatic nature of the undiluted ice lattice plays a significant role in the survival of the molecules. The disruption of the stiff $\pi$-$\pi$ interactions in the ice occurs regardless of the identity of the matrix component and appears to be the dominant factor influencing the variation in the destruction cross sections of aromatic molecules. Although the increase in the aromatic destruction cross section from undiluted to diluted ice is significant, it remains a factor of a few lower than that observed in a cyclohexane:water mixture (see App. \ref{app:cyclohexane}), indicating that the constraint of the fragmentation pathway preserving aromaticity continues to serve as an important factor in molecular resilience under UV exposure.
The significant variation in the cross sections of undiluted vs. diluted aromatic ices agrees with what has previously been reported by \citet{ruiterkamp2005quantitative}. \citet{ruiterkamp2005quantitative} reports a UV destruction cross section of 2.6$\times$10$^{-20}$ cm$^2$ for undiluted benzene and around 10$^{-18}$ cm$^2$ for mixed benzene ices. In particular \citet{ruiterkamp2005quantitative} reports an increase of the destruction cross section of mixed benzene ice in the order of H$_2$O$<$CO$<$CO$_2$ with respective values of 1.23$\times$10$^{-18}$ cm$^{2}$, 2.42$\times$10$^{-18}$ cm$^{2}$ and 4.38$\times$10$^{-18}$ cm$^{2}$. These values are consistent within a factor of a few with our experiments where we find cross sections of 3.9$\times$10$^{-18}$ cm$^{2}$, 4.3$\times$10$^{-18}$ cm$^{2}$ and  5.8$\times$10$^{-18}$ cm$^2$ respectively for benzene mixed with H$_2$O, CO, and \COO. The consistency of values across experiments performed on different setups and with varying procedures is quite reassuring, indicating that ice destruction cross sections are experimentally robust. 
Our upper limit of 10$^{-19}$ cm$^2$ for undiluted benzene also compares well with \citet{ruiterkamp2005quantitative} results. In addition, our experiments with substituted aromatic molecules, toluene and benzonitrile, show similar behavior, suggesting that the presence of an additional functional group does not significantly affect the photodestruction efficiency in the ice phase for aromatic molecules. Aromatic functionality appears to be the dominant factor governing UV stability in these systems, which was not obvious a priori given the potential influence of substituents on both electronic structure and interactions within the ice matrix. On the basis of these findings, we note that the decreased photoresistance observed in our diluted experiments can be more generally expected for the monosubstituted aromatic class of molecules as the dominant factor driving this change is the disruption of aromatic intermolecular interactions within the ice lattice.

The cross sections measured for diluted aromatic ices, while higher than those of pure ices, are however, still an order of magnitude smaller than the gas-phase photodestruction cross sections (Table \ref{tab_explist}). 
These differences can be partially attributed to a reduced cage effect in mixtures compared to pure aromatic ices where the $\pi$ interaction dominates but also implies that radical recombination is still important in these ice matrices as well. The ``stiffness" of the ice matrix, which reflects how strongly the molecules bind to each other, can be related to the desorption temperature of the constituent molecules of the matrix. Based on this reasoning, we would expect CO to induce the weakest cage effect, with water being the most caging of the three.
In other words, if direct recombination is the dominant factor we should hence expect to see photodestruction rates increase as H$_2$O$<$CO$_2$$<$CO, a trend that is not fully observed for any of our aromatic molecules (Fig. \ref{fig:fitall}). 

Chemical pathways promoted by the presence of the matrix must then also be considered. We begin with the case of a CO matrix, which should be the least chemically active of the three matrices we explored.
Dissociation of CO molecules does not occur within the photon energy range used in our experiments \citep{ruiterkamp2005quantitative}, and the formation of reactive excited CO (CO$^*$) molecules, capable of carbonylation of aromatic molecules \citep{Jamieson2006ApJS..163..184J,devine2022spin}, is not expected to be a significant reaction channel in the absence of catalysts. However, carbonylation of benzene has been observed in the presence of zeolites \citep{clingenpeel1997carbonylation}, which owing to their porous structure induce a ``concentration effect" for reactants within their pores \citep{Mouarrawis2018FrCh....6..623M}. Ices can, in some sense, have a similar ``concentration effect" making direct carbonylation a plausible route in icy environments. While the consumption of CO in our experiments is only a few percent, we speculate that carbonylation reaction might be playing a role for our benzene and toluene experiments in CO as some of the IR bands formed during irradiation could be assigned to aromatic aldehydes (Fig. \ref{fig:benzaldehyde} in App. \ref{app:productsfigandtable}). This can also help explain the $\sim$\,40\,$\%$ lower cross section measured for benzonitrile, as the deactivating effect of the cyano group can contribute to a lower efficiency in the carbonylation reaction.

Photolysis of CO$_2$ in the 120-170\,nm results in the formation of O($^1$D) + CO($^1$$\Sigma$) with an efficiency close to unity \citep{Zhu1990JChPh..92.2897Z,Slanger1971JChPh..54.1889S}. In experiments similar to ours, the insertion of O($^1$D) in the C-H bond has been proven to be a viable pathway for the functionalization of small hydrocarbons \citep{bergner2017methanol, bergner2019detection}, and it is likely to also play a role in combination with larger hydrocarbons. The formation of phenol of other alcohols is not obvious from the IR spectra 
presented in this work. However, dedicated experiments have shown that oxygen insertion reactions occur from the reaction of benzene and atomic oxygen (Piacentino \textit{in prep}). 

Water ice is also dissociated by our lamp, with the primary dissociation channel producing H and OH \citep{Yabushita2006JChPh.125m3406Y}. 
Radical addition reactions involving OH radicals with benzene, and toluene have been observed in the gas phase \citep{Zhan2018facile}. Notably, the gas-phase reaction with toluene produces higher product yields compared to benzene, due to the greater number of potential addition sites \citep{Atkinson1989formation}. Although we do not have direct confirmation of this chemistry in our data, we speculate that this reaction mechanism could be at play in the ice phase as well. In summary, based solely on matrix-induced reactivity, we would expect the apparent destruction cross section to follow the trend CO$<$H$_2$O$\leq$CO$_2$. In all cases, the CO matrices results in the lowest steady state yield. However the photodestruction cross sections only follow this trend in the case of benzonitrile (CO$<$H$_2$O$<$CO$_2$), and do not fully match the trends seen for benzene (H$_2$O$<$CO$<$CO$_2$) and toluene (CO$<$CO$_2$$<$H$_2$O).

Based on our results and reasoning, it is evident that neither the caging effect nor chemical reactivity alone can fully explain the observed trends; multiple factors, including direct recombination, ice diffusion, and matrix-molecule reactivity, must be acting simultaneously. Both the steady-state abundance and the destruction cross-section are therefore dependent on the combination of all these processes.
Although isolating each of these processes would require dedicated experiments, the overall impact of photon exposure on the abundance of the parent aromatic molecule in the ices can still be quantified and estimated. For molecules with similar photodestruction cross-sections in the gas phase, the corresponding cross-section in the ice phase is expected to be at least an order of magnitude lower in pure ices. This reduction is due to the onset of cage effects in the ices, which are more pronounced for molecules with stronger interactions and larger sizes. In mixed ices, however, the caging effect and the available chemistry are set by the matrix molecule. Together, these factors result in an increased organic photodestruction cross-section for mixed aromatic ices compared to pure. This represents a significant advancement over the gas-phase UV destruction cross-sections currently used in astrochemical models, providing a more comprehensive understanding of photon-driven chemistry in icy environments.

\subsection{Astrochemical consequences} \label{sec:astro}

The largest number of detections of aromatic molecules comes from cold interstellar clouds such as TMC-1 \citep{mcguire2018Sci...359..202M, mcguire2021detection}. Although benzene cannot be directly detected due to its lack of a dipole moment, the widespread detection of aromatic nitriles such as benzonitrile suggests that benzene is likely abundant in these environments. Benzene is thought to form in the gas phase under oxygen-poor conditions \citep{woods2002synthesis} and to rapidly freeze out onto dust grains due to its high sublimation temperature \citep{piacentino2024characterization}. The structure and composition of the resulting ice depend on the timing and conditions of its formation. 

Aromatic molecules may condense concurrently with water ice formation, becoming incorporated into the water ice, or may condense later on top of existing H$_2$O along with more volatile species like CO \citep{Pontoppidan08}. If aromatic molecules segregate within ices with low volatility such as water, they might form aggregates that exhibit properties more similar to pure ice rather than a mixture \citep{oberg2009quantification}. 
If, on the contrary, aromatics are condensed within volatile ices, subsequent thermal processing of the ice can lead to the desorption of the more volatile components like CO \citep{Pontoppidan08,Bergnericeinh2021ApJ...919...45B}, leaving behind a concentrated or even pure aromatic ice layer \citep{boogert2015observations}. This ice distillation has been proposed to explain the presence of pure CO$_2$ ices in protostellar envelopes \citep{van2006infrared}. Consequently, studying the survival of these molecules in both mixed and pure ices is essential for evaluating the potential inheritance of ices from the cloud to the disk stage.

The distribution of molecular species in disks is of particular relevance as it sets the
chemical inventory available for newly forming planets and planetesimals and, as such, the chemical complexity that can directly be inherited by these young objects. In the next few paragraphs we first discuss how the temperature profile of the disk affects the ice abundance of aromatic molecules, cyclohexane, and small C$_{(2-3)}$ hydrocarbons to define the disk region in which these molecules are expected to be in the ice phase. 
Next, we evaluate the lifetime of benzene and cycloexane ices in disks, by considering the ultraviolet (VUV) flux that is expected at different radii and disk heights. By doing this we also define a region of the disk in which aromatic chemistry is expected to be the most significant.

The disk temperature profile as well as the VUV flux are modelled from the DALI code \citep{Bruderer2012A&A...541A..91B, Bruderer2013prpl.conf2B054B}. DALI derives a 2D temperature, radiation field, and chemical abundance given a user-defined physical structure, stellar spectra, external radiation field, and initial chemical abundances. It calculates the gas temperature iteratively as chemistry evolves and provides excess heating and cooling terms beyond what is provided by the central star. In this example, we model a typical disk around a K star, with a total disk mass of 1\,$\%$ of the stellar mass (0.6 Msun), gas-to-dust ratio of 1000, and a stellar C/O ratio (0.47). 

\begin{figure} [h!]
\centering
\includegraphics[width=1.1\columnwidth]{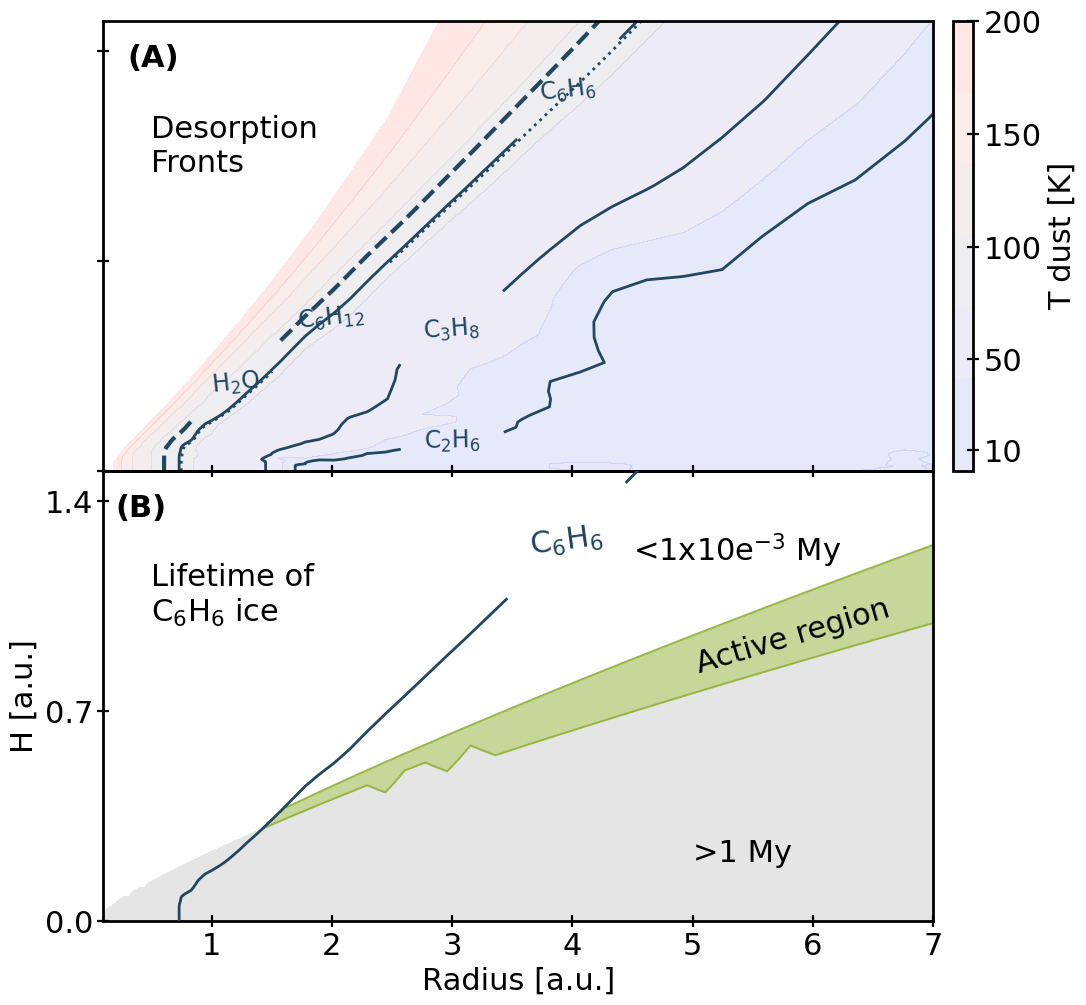}
\caption{(A) Temperature distribution in disks and snowline location for pure benzene, cyclohexane (dotted for clarity), propane and ethane ice. (B) Identification of the active zone for C$_6$H$_6$ ice in a disk model.}
\label{fig:disk}
\end{figure}

The disk temperature profile and the molecules binding energy are the main parameter to determine whether a molecule is in the ice-phase at a given disk radius and height. Fig. \ref{fig:disk} (A), shows the desorption fronts, or snowlines, of C$_2$H$_6$, C$_3$H$_8$, C$_6$H$_6$ and C$_6$H$_{12}$ overlayed with disk temperature profile from the thermo-chemical model.
We determine the aromatic molecules desorption fronts by calculating the equivalence point of the freeze out and desoprtion rate. The freeze out rate is calculated using the procedure described by \citet{Du2014ApJ...792....2D} as follow,

\begin{equation} \label{eq: desfront}
    k_F = S\sigma v_T n_d
\end{equation}

Where S is the sticking coefficient set to unity, $\sigma$=1x10$^{-5}$ cm$^2$ is the geometrical cross section of the dust grain, n$_d$ is the number density of dust grains, and \textit{v$_T$} is the thermal velocity of the molecule with mass \textit{m}.

\begin{equation} \label{eq: themvel}
    v_{T} = \sqrt{\frac{8*k_b*T}{\pi*m}}
\end{equation}

Similarly, the desorption rate is calculated using, 

\begin{equation} \label{eq: desrate}
    k_{d} = \nu_0  e^\frac{-BE}{T}
\end{equation}

Where BE is the binding energy of each molecule, and $\nu$$_i$ is the associated pre-exponential factor. In our calculation we use experimental BE and $\nu$$_i$ values compiled from \cite{piacentino2024characterization} for benzene, and \citet{behmard2019desorption} for ethane and propane. Since desorption parameters for cyclohexane were not available in the literature, we derived them experimentally using a procedure similar to that described in \citet{piacentino2024characterization}, as detailed in App. \ref{app:cyclohexane}. 
As shown in Fig. \ref{fig:disk} (A), the snowlines of the different molecules vary significantly leading to different distribution between gas and ice-phase of the hydrocarbons at different points within the disk.  
Given the relative positions of the snow lines for the hydrocarbons in our model and the water snowline (Fig. \ref{fig:disk}, A), we expect the abundance of aromatic ice as well as that of cyclohexane to remain unaffected by thermal processing, while smaller hydrocarbons will desorb into the gas-phase across much of the disk radius.

Consequently, the interaction with UV photons is likely the dominant process affecting the abundance of both cyclohexane and aromatic molecules in disks. Based on the VUV destruction cross sections calculated in this work for benzene ice and the values of photon fluxes obtained from the 2D thermochemical disk model, it is possible to identify three regions in the disk that correspond to different regimes of icy aromatic molecular lifetimes. Although we opted to use the destruction cross section of pure benzene, the similarity in cross section values among aromatic molecules means this approach can be extended to other molecules. The same model can also be applied to calculate the lifetimes of aromatics in different ice environments. As a result, our calculation is influenced by the uncertainty in ice composition, which is larger than the uncertainty associated with the cross sections themselves and is closer to 20$\%$.

At higher disk heights the lifetime of icy aromatic molecules is lower than 1000 years. In this region, we should expect the aromatic ice abundance to be significantly reduced by photon exposure. Conversely, closer to the disk midplane, photodissociation of aromatic molecules is expected to be minimal due to shielding resulting in the aromatic molecules lifetime to be longer than 1 My. 
Finally, at intermediate disk heights (0.3 to 1.1 a.u., green shaded area in Fig. \ref{fig:disk}, B), we find a region, starting at radii $\sim$\,1.5 a.u. from the central star, in which the photon flux is low enough to conserve some of the aromatic inventory, while being sufficiently high to induce photochemistry in the aromatic substrates. This has implications for the aromatic inventory found in the ices of protoplanetary disks, as the processes occurring in this region could increase the complexity of the aromatic species available during planet formation.
Given the similarity in desorption kinetics between benzene and cyclohexane (Fig. \ref{fig:disk}), but the notably different measured destruction cross sections in both pure (Fig. \ref{fig:allpure}) and mixed ices (Fig. \ref{fig: cyclohexane decay} in appendix \ref{app:cyclohexane}), we compare the lifetime of these two molecules in terms of unitless scale heights based on our model. The active region spans roughly 2-3 scale heights, depending on radius, with cyclohexane only surviving closer to the midplane than benzene in both pure and diluted ices. However, this difference in survival depth between the two molecules decreases when the molecules are embedded in water ice. In either case, aromatic molecules, being more photoresistant, persist at higher disk heights which, over time, should increase the aromatic/aliphatic ratio in disk ices. This is especially true when considering that these complex aromatic molecules do not remain fixed at a single disk height throughout disk's lifetime but rather participate in the vertical mixing that continuously redistribute dust grains and ices between the surface and midplane \citep{ciesla2010residence}. In time, this dynamic circulation can enrich the midplane with functionalized aromatic molecules formed higher up in the disk, increasing both the aromaticity and the chemical complexity available in the planet-forming region.

\begin{figure}[t!]
    \centering
    \includegraphics[width=\linewidth]{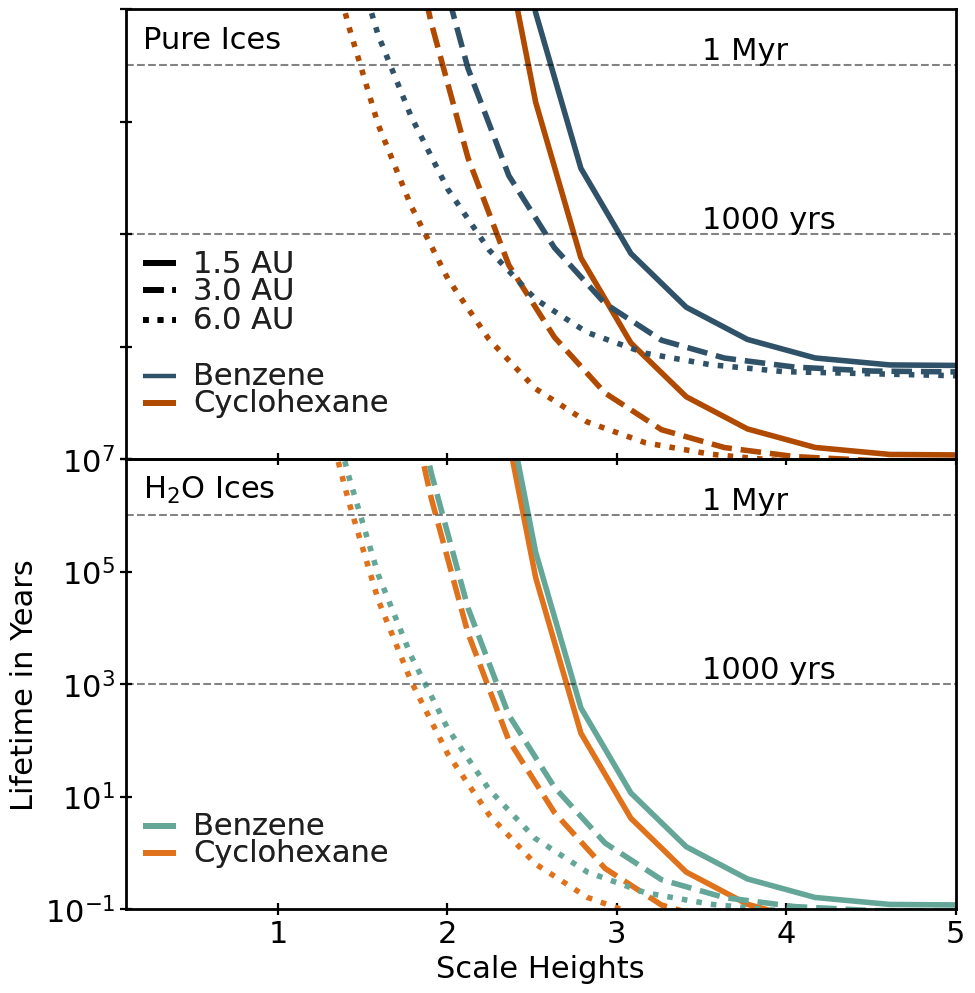}
    \caption{The life time of C$_{6}$H$_{6}$ and C$_{6}$H$_{12}$ ices as a function of scale heights at 1.5, 3, and 6 a.u. The black dashed lines indicate 1 Myr and 1000 years. Between the dashed gray indicates at what scale height each molecule will be processed on a typical disk timescale.}
    \label{fig:scaleheights}
\end{figure}

\section{Conclusion} \label{sec:conc}
We studied the behavior of three aromatic molecules ice in the presence of VUV radiation and in combination with three matrix constituents CO, \COO, and H$_2$O. We found that:
\begin{itemize}
    \item In undiluted ices, recombination is more efficient in aromatic than aliphatic ices, resulting in significantly lower destruction cross sections. This indicates that aromaticity inherently stabilizes ices against UV processing in astrophysical environments.
    \item No discernible substituent effects can be observed in pure aromatic ices. The ices of aromatic molecules are resistant to VUV exposure regardless of the ring functionalization, as demonstrated by consistent destruction cross sections across a series of substituted aromatics, including benzonitrile and alkylbenzenes up to butylbenzene. This reinforces the conclusion that aromaticity itself is responsible for the high photostability of these ices. The exact threshold at which substituent size begins to affect the aromatic UV resistance requires further investigation. 
    
    \item When diluted, the destruction cross section of the aromatic ices increases significantly, thanks to a weaker caging effects of the matrix constituents as well as possible chemical attacks from matrix fragments. Yet the dissociation cross sections remains significantly lower than the gas-phase counterpart.
    \item The nature of the matrix appears to impact the chemistry available to the aromatic molecules and dedicated experiments are needed to fully understand the effects.
    \item In disks, the lifetime of icy aromatic molecules depends on the local VUV flux, defining a layer above the midplane where aromatic photochemistry is most active. Over time, vertical mixing causes these molecules to sink toward the midplane, enriching the planet-forming region with both aromaticity and chemical complexity.    
\end{itemize}

All the IR experimental data are available at 10.5281/zenodo.15278527 and 10.5281/zenodo.15247573.

This work was supported by a grant from the Simons Foundation 686302, KÖ. E.L.P. thanks Alexandra McKinnon for helpful discussions related to this work.

\bibliography{sample631}{}
\bibliographystyle{aasjournal}

\newpage
\appendix{}\label{sec:appendix}

\section{Absorption cross sections of the matrix components}\label{app:lamp}

\begin{figure}[thb]
\centering
\includegraphics[width=0.6\textwidth]{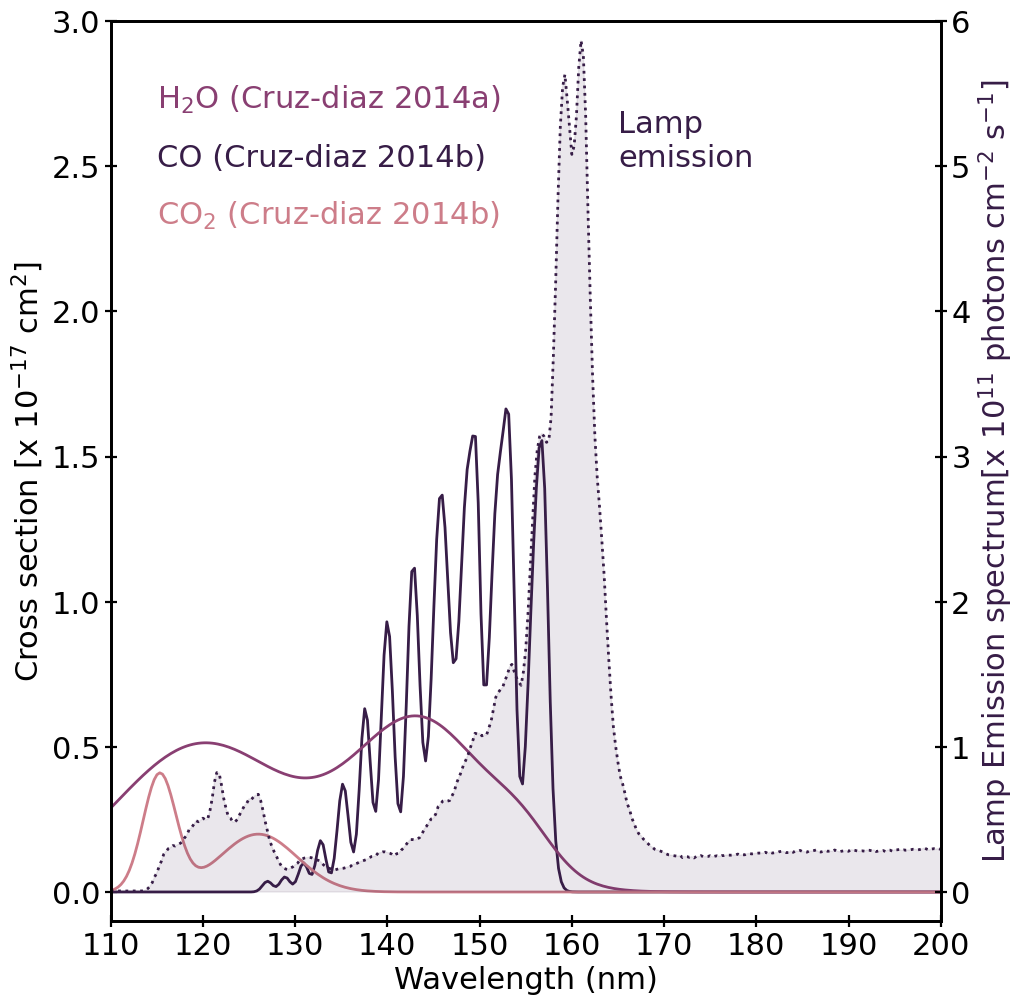}
\caption{Photon absorption profiles compared to the laboratory lamp emission spectra. }
\label{figapp: lamp}
\end{figure}

The emission profile of the lamp used in our experiment is shown in Fig. \ref{figapp: lamp} (shaded region), we overlap the ice phase photon absorption profile of the matrix constituent. These values have previously been reported in \citet{pCruz-Diaz2014AA...562A.119C} and \citet{nonpCruz-Diaz2014AA...562A.120C}. The absorption profile of CO is, among the molecules used as matrices, the one that most overlaps with our lamp emission; for this reason, CO was chosen as the reference scenario for which the coverage limits on the optically thin regime for mixtures as well as the uncertainty in our reported cross sections are evaluated. 

The fraction of photon sequestered by the matrix can also be calculated analytically using Eq. \ref{eqatten}   

\begin{equation} \label{eqatten}
I_t(\lambda) = I_0(\lambda) \cdot e^{-\sigma(\lambda) \cdot N_x}
\end{equation}

Where $I_t(\lambda)$ and $I_0(\lambda)$ are the transmitted and incident intensities,  $N_x$ is the molecular column density (cm$^{-2}$), and $\sigma$($\lambda$) is the absorption cross sections of the molecules constituting the matrices.
Using this equation, and the absorption cross section in Table \ref{table_combined} and in \citet{nonpCruz-Diaz2014AA...562A.120C} for CO, we find that for total ice thicknesses of 185 ML (see \S\ref{subsec:unce}) the photon sequestration by CO approaches 15\,$\%$ at 160\,nm and it is $\sim$\,56\,$\%$ on average in the 120-160\,nm spectral window range.

\newpage
\section{Variation of the photodestruction of benzene with ice coverage} \label{app:thickandtempbenzenepure}

\begin{figure}[thb]
\centering
\includegraphics[width=0.45\textwidth]{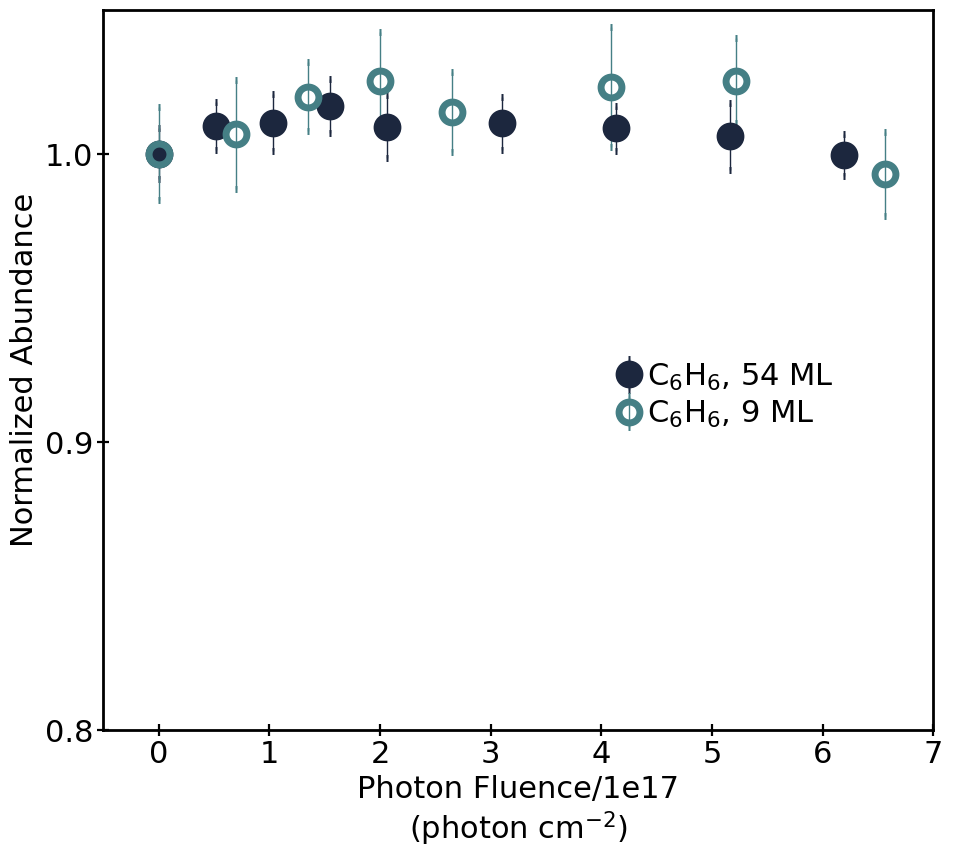}
\caption{Variation in the abundance of benzene in the ice vs. fluence for two ice coverages.}
\label{fig:benzene thickness}
\end{figure}

Photodestruction of benzene has been studied in undiluted ices at two different ice coverages (Fig. \ref{fig:benzene thickness}) to ensure that all experiments were in the optically thin regime. Our experiments show that is the case, as there is no sensible variation in the destroyed fraction measured in a 9 and a 55 ML ices. 

\newpage
\section{Cyclohexane, survival in water mixture and binding energy}\label{app:cyclohexane}

\begin{figure} [h!]
\centering
\includegraphics[width=0.5\columnwidth]{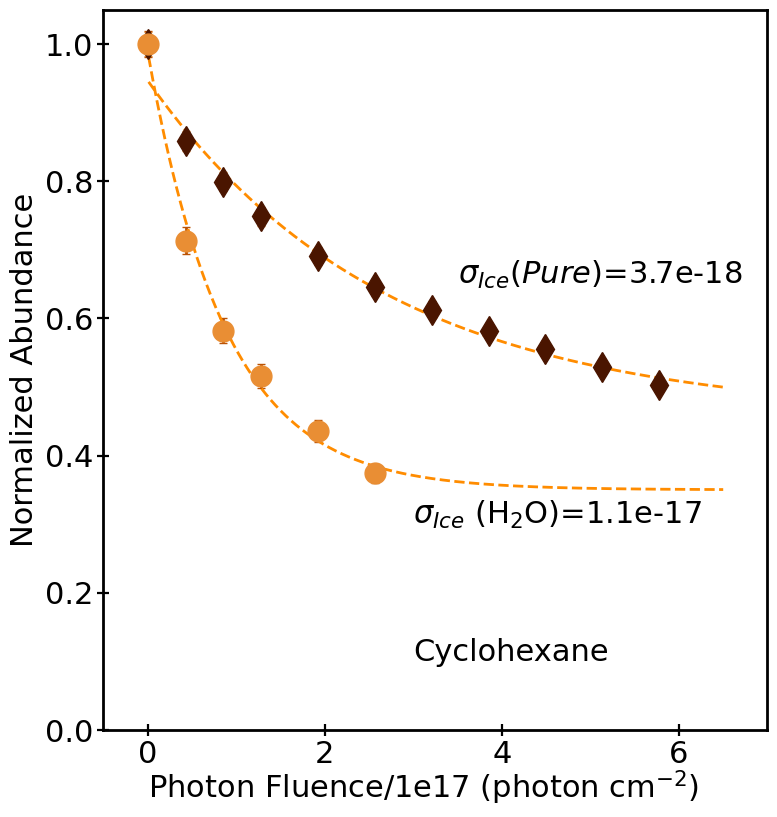}
\caption{Destruction cross section of a  C$_6$H$_{12}$:H$_2$O ice compared to the undiluted cyclohexane ice.}
\label{fig: cyclohexane decay}
\end{figure}

An additional experiment was run to study the destruction cross section of cyclohexane in a water matrix. The procedure used to carry out the experiment and evaluate the cross section is equivalent to that described in the main text. The resulting ice had a C$_6$H$_{12}$:H$_2$O mixing ratio of 1:10, with a total ice coverage of 121 ML. The fitted destruction cross section, compared to that of pure cyclohexane ice, is shown in Fig. \ref{fig: cyclohexane decay} with a $\sigma_{Ice}$ of 1.1\,$\times\,$10$^{-17}$ cm$^2$ and a steady state yield of 0.35. We found that the presence of water enhances the destruction of cyclohexane compared to the undiluted case; however, the enhancement is not as pronounced as what is observed for aromatic molecules (Fig. \ref{fig:fitall}). This provides additional evidence for the importance of aromaticity as a protective factor contributing to molecular survival under UV exposure. 

\begin{figure} [h!]
\centering
\includegraphics[width=0.7\columnwidth]{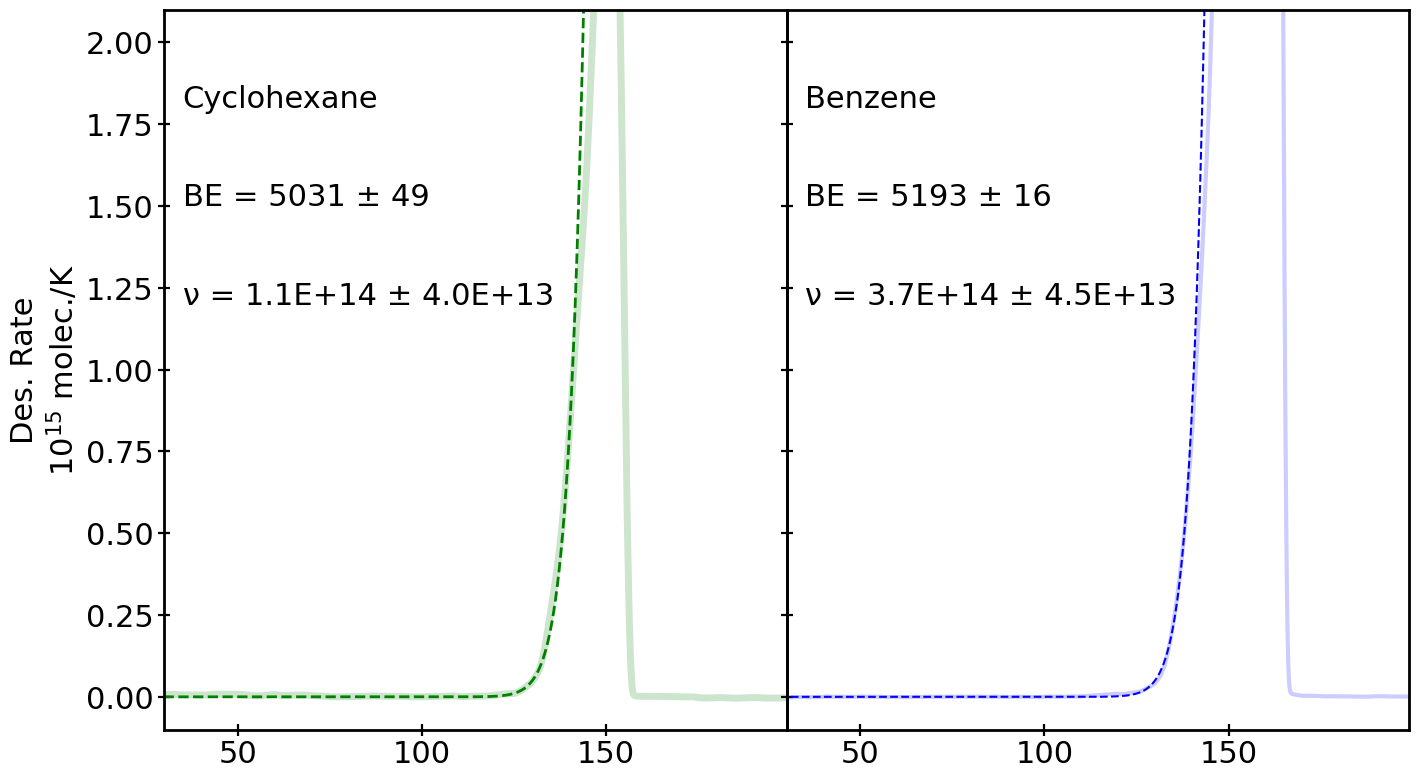}
\caption{TPD of pure cyclohexane ice, dashed line is the best fit obtained using the Polanyi–Wigner equation, best fit parameters are reported in the figure.}
\label{fig: be cyclohexane}
\end{figure}

As an input for the model described in \S\ref{sec:astro}, we needed to estimate the binding energy and the attempt frequency of cyclohexane ice. We chose to determine this experimentally, following the method reported in \citet{piacentino2024characterization}. Specifically, we performed a pure cyclohexane TPD experiment, using a desorption rate of 2\,K/min from an 34\,ML ice. The resulting TPD profile was fitted using the Polanyi–Wigner equation to obtain both the attempt frequency and the binding energy of cyclohexane, as detailed in \citet{piacentino2024characterization}. The fit result is shown in the left panel of Fig. \ref{fig: be cyclohexane}, in comparison with the binding energy fit of a benzene ice (right panel)
We note that the derived binding energy values may carry a larger uncertainty than the nominal fit uncertainty reported in Fig. \ref{fig: be cyclohexane}, as they were obtained from a single experimental run.

\newpage
\section{Irradiated Ice IR Spectra}\label{app:productsfigandtable}
\begin{figure} [h!]
\centering
\includegraphics[width=0.8\textwidth]{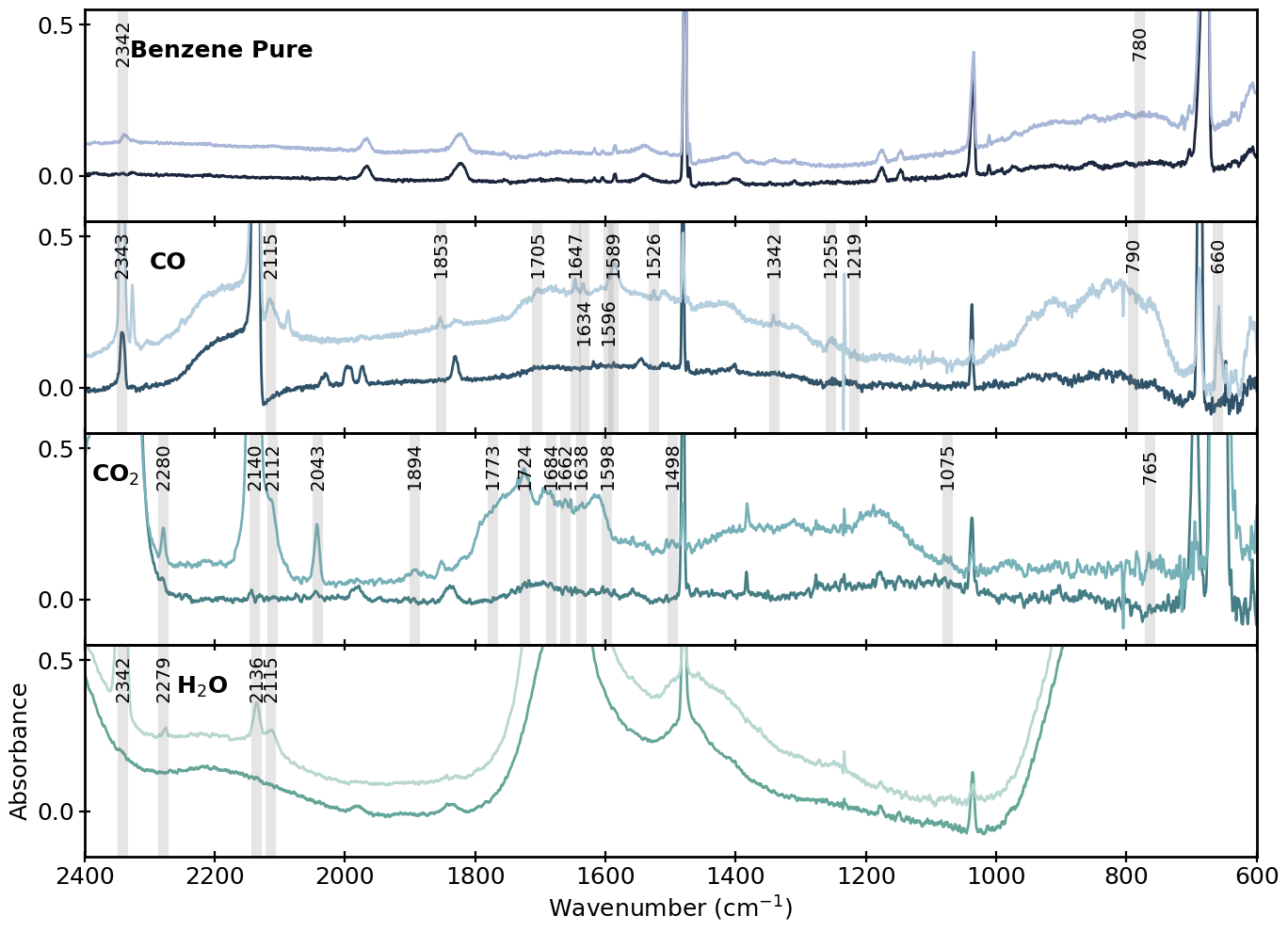}
\caption{Spectra of benzene pure and mixed ices before (darker line) and after (lighter line) photon irradiation. Intensities are normalized to better evidence the product bands which are labeled in the plots.}
\label{fig: benzene producs}
\end{figure}

\begin{figure} [h!]
\centering
\includegraphics[width=0.8\textwidth]{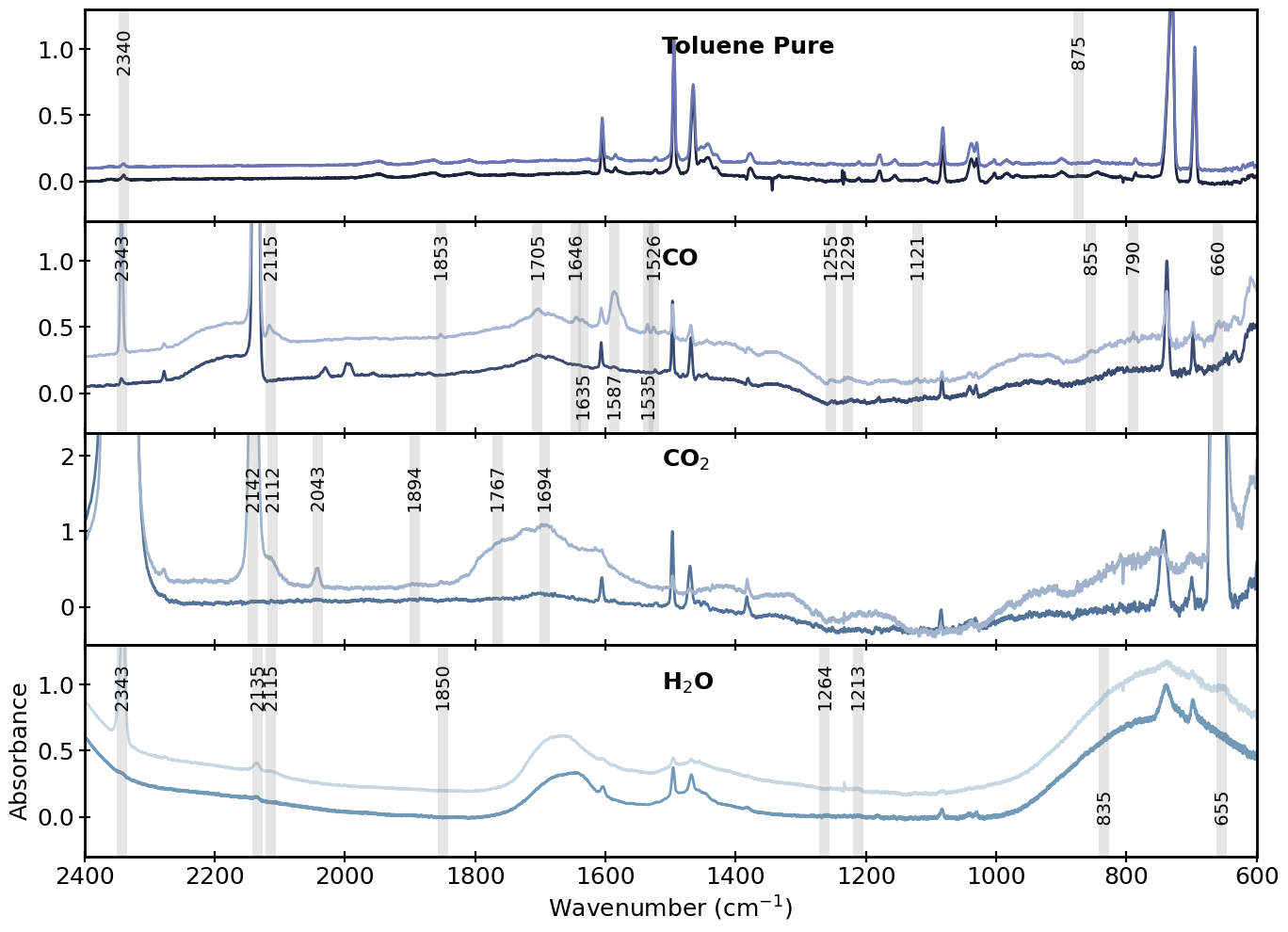}
\caption{Spectra of toluene pure and mixed ices before (darker line) and after (lighter line) photon irradiation. Intensities are normalized to better evidence the product bands which are labeled in the plots.}
\label{fig: toluene products}
\end{figure}

\begin{figure} [h!]
\centering
\includegraphics[width=0.8\textwidth]{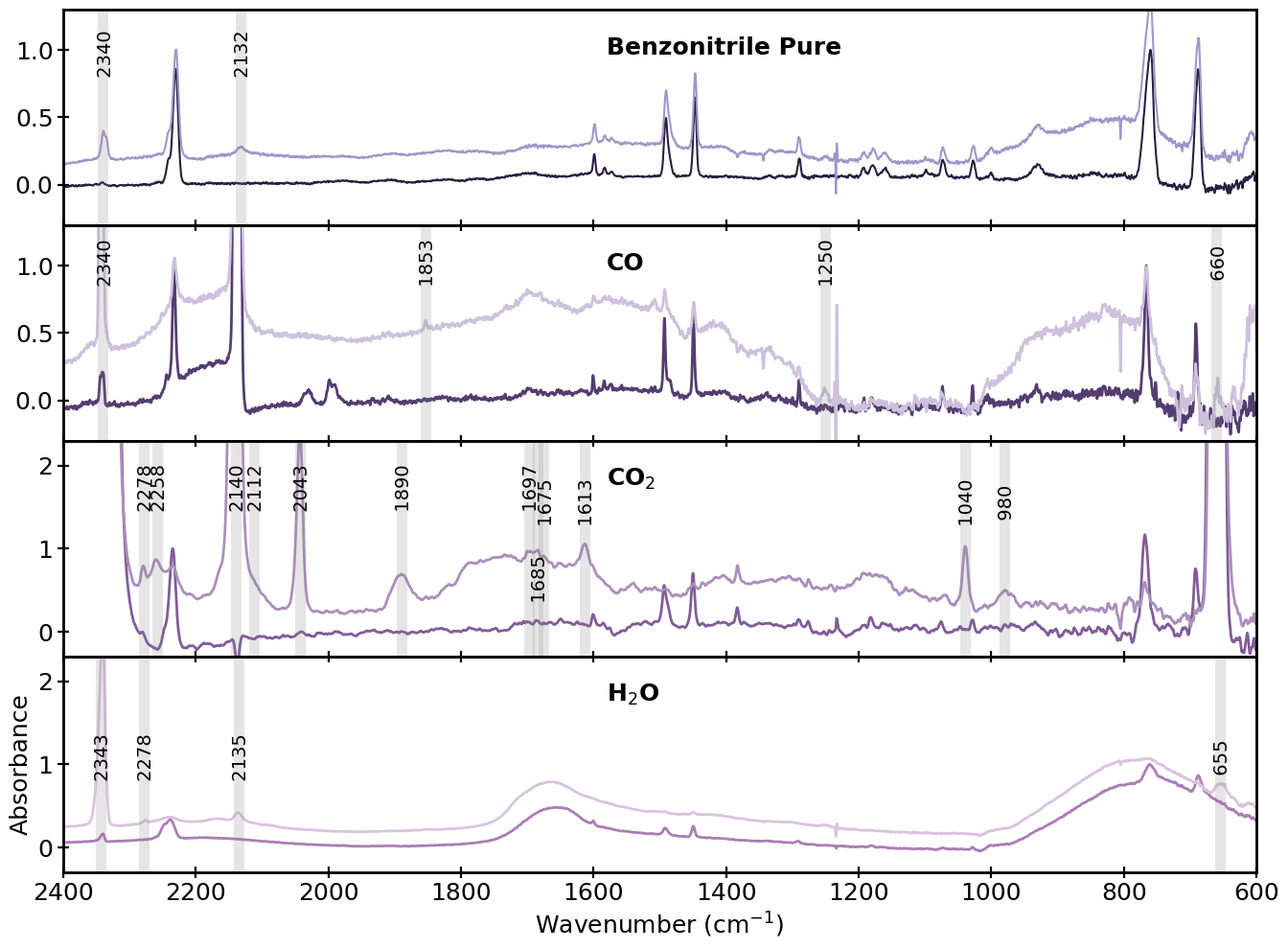}
\caption{Spectra of benzonitrile pure and mixed ices before (darker line) and after (lighter line) photon irradiation. Intensities are normalized to better evidence the product bands which are labeled in the plots.}
\label{fig: benzonitrile products}
\end{figure}

\begin{figure} [h!]
\centering
\includegraphics[width=0.8\textwidth]{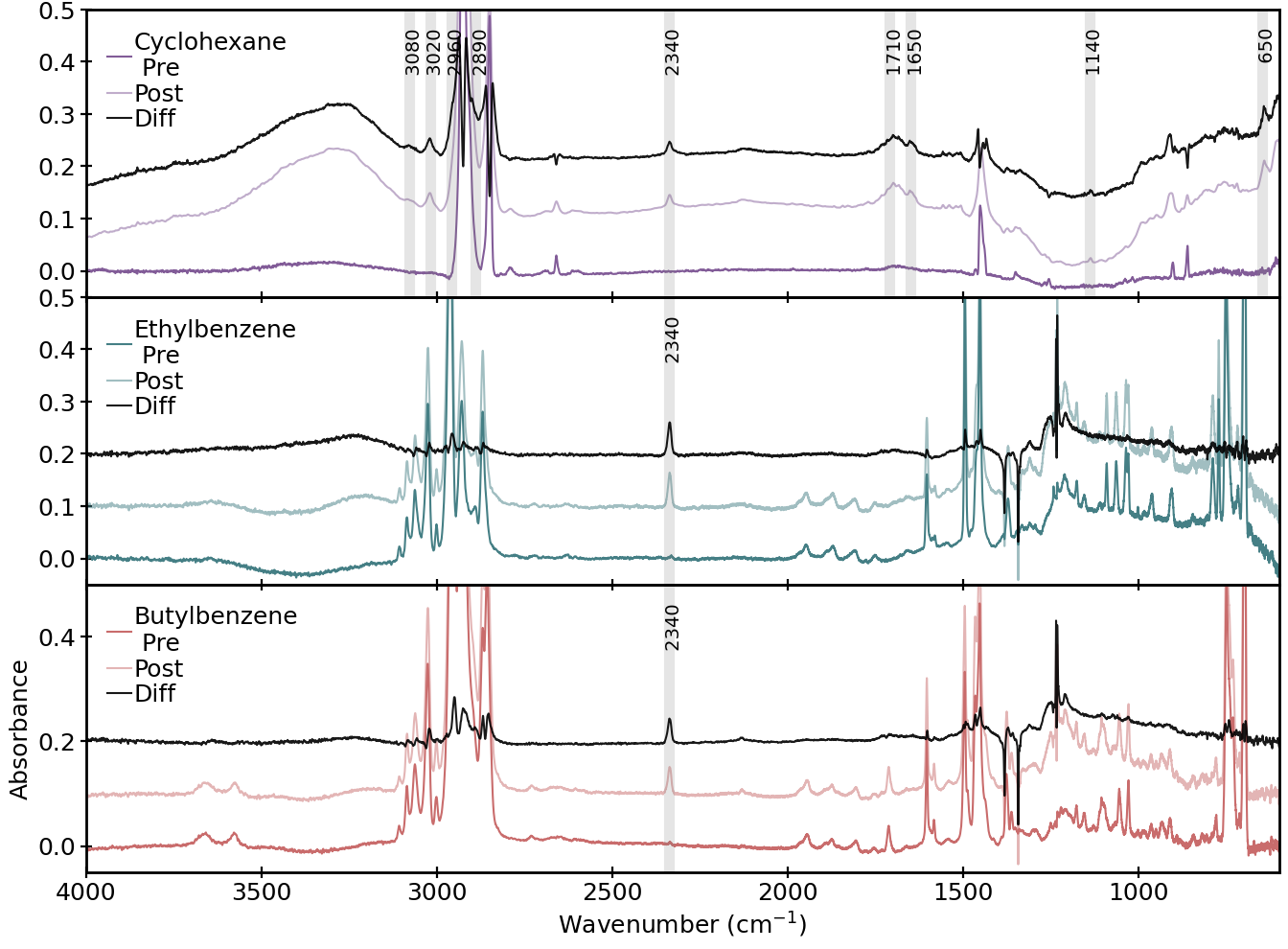}
\caption{Spectra of pure cyclohexane, ethylbenzene, and butylbenzene ices, before (darker line) and after (lighter line) photon irradiation. The black lines show the difference spectra. Intensities are normalized to highlight product bands, which are labeled. New peaks are shaded in gray. Variations in the aromatic molecules band intensities during irradiation are attributed to changes in band strength caused by photon exposure.}
\label{fig: add products}
\end{figure}

\begin{figure} [h!]
\centering
\includegraphics[width=0.45\columnwidth]{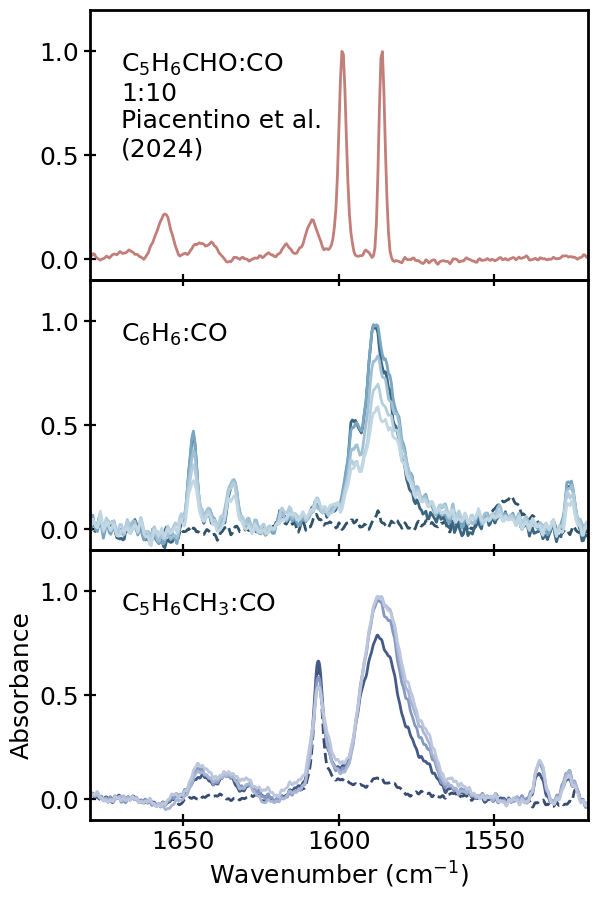}
\caption{The IR spectra of benzaldehyde:CO ice, reproduced from 
\citet{piacentino2024characterization} (Top panel), in comparison with the spectra of benzene:CO (middle panel) and toluene:CO (bottom panel) before (dashed lines) during UV irradiation (solid lines). }
\label{fig:benzaldehyde}
\end{figure}

In Fig. \ref{fig: benzene producs}, \ref{fig: toluene products}, and \ref{fig: benzonitrile products} we report the bands formed during the irradiation of each experiment that uses benzene, toluene and benzonitrile respectively. We do not aim to provide product identification, as accurate band assignment would require a detailed study of each experimental condition, which falls beyond the scope of this work. 
However, we point out that the irradiation of the benzene:CO and toluene:CO ices produces IR bands in the 1500-1700 \cmu range, which can be attributed to the formation of an aldehyde functionality. The product bands are depicted in Fig. \ref{fig:benzaldehyde} and are directly compared to the spectra of a 1:10 benzaldehyde:CO ice reproduced from \citet{piacentino2024characterization}. Although product assignment from our current data can only be speculative, we point out the strong similarities between the newly formed bands and the vibrational modes of benzaldehyde ice. Fig. \ref{fig: add products}, shows the pre and post irradiation spectra for the cyclohexane, ethylbenzene and butylbenzene pure ices. Formation of new bands is observed during photon irradiation of cyclohexane. However, ethylbenzene and butylbenzene spectra shows similar behavior to what observed in the other pure aromatic ices in this study and no significant spectral variation is induced by photon exposure.

\end{document}